\documentclass[lettersize,journal]{IEEEtran}
\usepackage{amsmath,amsfonts}
\usepackage{algorithmic}
\usepackage{algorithm}
\usepackage{array}
\usepackage[caption=false,font=normalsize,labelfont=sf,textfont=sf]{subfig}
\usepackage{textcomp}
\usepackage{stfloats}
\usepackage{url}
\usepackage{verbatim}
\usepackage{cite}
\usepackage{graphicx}
\usepackage{xcolor}
\usepackage{booktabs}
\usepackage{multirow}

\newtheorem{definition}{Definition}

\newboolean{showcomments}
\setboolean{showcomments}{false} 
\ifthenelse{\boolean{showcomments}}
{\newcommand{\nb}[2]{
		\fcolorbox{black}{yellow}{\bfseries\sffamily\scriptsize#1}
		{\sf\small$\blacktriangleright$\textit{#2}$\blacktriangleleft$}
	}
	
}
{\newcommand{\nb}[2]{}
	
}

\makeatletter
\newcommand{\figcaption}[1]{\def\@captype{figure}\caption{#1}}
\newcommand{\tblcaption}[1]{\def\@captype{table}\caption{#1}}
\makeatother

\def\BibTeX{{\rm B\kern-.05em{\sc i\kern-.025em b}\kern-.08em
    T\kern-.1667em\lower.7ex\hbox{E}\kern-.125emX}}
\begin{document}
\bstctlcite{IEEEexample:BSTcontrol}

\newcommand\jialong[1]{\nb{Jialong}{#1}}
\newcommand\sherry[1]{\nb{Sherry}{#1}}
\newcommand\kenji[1]{\nb{Kenji}{#1}}

\newcommand{\methodname}{\textsc{EndWeaver}}
\newcommand{\high}{method}
\newcommand{\High}{Method}
\newcommand{\low}{primitive task}
\newcommand{\Low}{Primitive task}
\newcommand{\LOW}{Primitive Task}

\title{Automated Smoke Testing for Multi-ending Game Quests via Large Language Models}

\author{
    Jinyu Cai\textsuperscript{1},
    Mingyue Zhang\textsuperscript{2},
    Nianyu Li\textsuperscript{3},
    Jinglue Xu\textsuperscript{4},
    Tianshui Li\textsuperscript{5},
    Jialong Li\textsuperscript{1,*},
    Hitoshi Iba\textsuperscript{5},
    and Kenji Tei\textsuperscript{6}
    \thanks{\textsuperscript{1}Jinyu Cai and Jialong Li are with Waseda University, Tokyo, Japan.}
    \thanks{\textsuperscript{2}Mingyue Zhang is with Southwest University, Chongqing, China.}
    \thanks{\textsuperscript{3}Nianyu Li is with ZGC Laboratory, Beijing, China.}
    \thanks{\textsuperscript{4}Jinglue Xu is with Sakana AI, Tokyo, Japan.}
    \thanks{\textsuperscript{5}Tianshui Li and Hitoshi Iba are with the University of Tokyo, Tokyo, Japan.}
    \thanks{\textsuperscript{6}Kenji Tei is with the Institute of Science Tokyo, Tokyo, Japan.}
    \thanks{\textsuperscript{*}Corresponding author: lijialong@fuji.waseda.jp.}
    \thanks{Accepted for publication in IEEE Transactions on Games. DOI: 10.1109/TG.2026.3720476. \copyright\ 2026 IEEE. Personal use of this material is permitted. Permission from IEEE must be obtained for all other uses, in any current or future media, including reprinting/republishing this material for advertising or promotional purposes, creating new collective works, for resale or redistribution to servers or lists, or reuse of any copyrighted component of this work in other works.}
}

\maketitle


\begin{abstract}
Modern game quests are increasingly trending toward high freedom and multiple endings, making their design and testing more complex. {\color{black}Smoke testing in games aims to rapidly validate quest functionality, but existing automated testing approaches face critical limitations when handling quests with branching narratives, and multiple possible endings. These challenges result in insufficient exploration of endings, redundant action sequences, and prohibitively high interaction costs.}
To address these challenges, we propose \methodname{}, an automated smoke testing framework specifically designed for multi-ending game tasks. \methodname{} leverages LLMs to interpret quest descriptions and game environments, generating high-level \high{}s to guide global exploration, complemented by the generated \low{} sequences for fine-grained planning. By combining this multi-level planning structure with the dynamic construction of state-transition graphs, \methodname{} ensures comprehensive coverage of all possible endings and enables efficient testing.
We evaluated \methodname{} on {\color{black}four game environments---Mini RPG, Overcooked Plus, Minecraft, and PySC2} against baseline methods. In {\color{black}all environments, \methodname{} achieved stable quest completion with high reliability}, significantly outperforming the baselines. In terms of testing cost, \methodname{} required only a few thousand steps, while baseline methods required millions of steps, all while ensuring comprehensive coverage of all possible quest endings.
Ablation studies further demonstrated the critical contributions of each framework component, validating the effectiveness and robustness of our approach.

\end{abstract}

\begin{IEEEkeywords}
Smoke Testing, Large Language Model, Role-playing Game
\end{IEEEkeywords}

\section{Introduction}
{\color{black} 
In recent years, the gaming industry has expanded rapidly, with players demanding richer narrative freedom and greater content diversity~\cite{vartinen2022generating}. Quests, as the core vehicle of player engagement, now frequently adopt multi-ending structures and highly flexible branching mechanisms~\cite{smith2011situating}. While these features enhance player immersion, they also impose substantial complexity on quest design and verification. At the same time, live-service operations—characterized by seasonal updates, recurring events, and balance patches—have accelerated development cycles, with some studios releasing multiple internal builds per day and updating hundreds of quests weekly~\cite{yuechen2020regression}. The combination of game complexity and frequent iteration cycles makes testing increasingly essential for maintaining both development velocity and consistent player experience quality.

As a built-in step in quest development, smoke testing~\cite{mcconnell1996daily} is used to verify whether newly implemented content can be executed properly~\cite{9787844}. Unlike traditional testing methods that aim for comprehensive bug detection, smoke testing focuses on providing early signals of executability—ensuring that newly created quests or levels can be played without critical issues~\cite{jie2024alanguage}. 
This type of testing—verifying the correctness of core gameplay flows at an early development stage—is particularly critical in high-frequency development environments \cite{Herbold2022}. First, it provides game designers with rapid and explicit feedback to confirm the correctness of each iteration and avoid misunderstandings between game design and development teams~\cite{politowski2021survey}. Second, it prevents full-scale testing from proceeding when core functionality is broken, thereby avoiding the costly waste of testing resources \cite{memon2003dart}.
However, as multi-ending quests become increasingly complex and updates occur more frequently, manual smoke testing faces significant challenges in terms of cost, consistency, and coverage—underscoring the need for automation.

Automated playtesting in games has been extensively studied. It initially emerged from model-based methodologies that aimed to simulate gameplay behavior through structured representations, including predefined scenario scripting~\cite{cho2010scenario}, replaying recorded user interactions~\cite{ostrowski2013automated}, and generating test cases from formal models such as UML statecharts~\cite{iftikhar2015automated} and Petri nets that describe game logic~\cite{becares2017approach}.
Subsequently, to better scale with the complexity of modern games, research has increasingly shifted toward agent-based methods, particularly those leveraging reinforcement learning (RL) agents capable of interacting with game environments for large-scale state-space exploration~\cite{bergdahl2020augmenting,sestini2024ccpt,9619048}. For example, Zheng et al.\cite{zheng2019wuji} combine deep reinforcement learning with evolutionary algorithms to balance “winning” (progressing through the game) and “exploration” (covering state space), thereby enhancing targeted bug discovery.
More recently, to better reflect the diverse preferences of human players, some studies have proposed persona-centered approaches, in which agents simulate human play styles to make testing more representative of real gameplay. These include methods that employ procedurally generated personas\cite{holmgaard2018automated} or player styles learned from expert demonstrations~\cite{ahlberg2023generating,ariyurek2019automated}, enabling test agents to capture a broader range of player behaviors.

However, the above approaches are not directly suited for smoke testing, which emphasizes rapid executability verification in multi-ending quests rather than exhaustive state coverage and bug discovery. Their utility, in the purpose of smoke testing, is limited by three key issues: 
(i) existing approaches often require millions or even billions of environment interactions to ensure convergence, and when encountering new quests, they usually need retraining or parameter adjustments, leading to prohibitively high interaction costs;
(ii) these approaches, especially RL-based methods, usually rely on reward functions for optimization, which not only suffer from sparse reward distributions that hinder effective learning, but also tend to converge to a single path in multi-ending quests, resulting in insufficient exploration;
(iii) such approaches frequently generate test sequences with redundant or inefficient actions, making it difficult to produce compact and high-quality optimal paths.

To address these issues, we propose a novel automated smoke testing framework, named \methodname{}, with three core ideas: (i) leveraging large language models (LLMs) for their text understanding and broad knowledge to support reasoning over game quests; (ii) multi-level planning to ensure comprehensive exploration and efficient testing; and (iii) combining state-action transition graphs to eliminate redundant actions.
For (i), \methodname{} leverages LLMs to interpret quest descriptions and game environments, thereby enabling semantic reasoning across diverse quests and reducing excessive environment interactions.
For (ii), \methodname{} adopts a multi-level planning approach, decomposing quests into high-level abstract \high{} and low-level specific \low{}, thereby enabling comprehensive exploration to cover all possible paths in multi-ending quests. 
For (iii), \methodname{} incorporates state-action transition graphs, which record action sequences and state transitions to eliminate redundant nodes and actions, thereby achieving the optimal paths.
}

The main contributions of this paper are as follows: 
\begin{itemize} 
\item We introduce a novel framework, \methodname{}, tailored for smoke testing multi-ending game quests. This framework leverages LLMs to understand and reason about the game quest, and utilizes hierarchical reasoning along with graph structure construction to enable comprehensive and efficient exploration of multiple endings.
\item To address the unique challenges of testing multi-ending game quests, \methodname{} integrates three key components: (i) leveraging LLMs to interpret quest descriptions and reason about diverse quest structures, reducing excessive environment interactions; (ii) combining \high{} generation with \low{} sequence planning to balance broad path exploration with precise quest execution; and
(iii) designing a dynamic state-action transition graph to enable efficient path coverage, redundancy reduction, and critical path identification.
\item We conduct comprehensive experimental evaluations, demonstrating the superior performance of our framework in terms of efficiency, path diversity, and computational cost compared to traditional baseline methods. Furthermore, we perform a thorough ablation study, which evaluates the contribution of each key component (i.e., LLM, multi-level planning, and the state-action transition graph).
\end{itemize}

The rest of the paper is organized as follows: Section~\ref{sec:relatedwork} reviews related work on automated game testing and studies on LLMs in the gaming field. 
Section~\ref{sec:problem} defines the notation and problem for the smoke testing of multiple-ending quests. 
Section~\ref{sec:proposal} explains the technical details of the \methodname{}.
Section~\ref{sec:evaluation} provides a comprehensive evaluation of the \methodname{} through {\color{black}four} games.
Finally, Section~\ref{sec:conclusion} concludes the paper and explores potential directions for future research.

\section{Related Work}
\label{sec:relatedwork}
\subsection{Game Testing}
Game testing plays a critical role in the lifecycle of game software, aiming to ensure the stability, completeness, and overall user experience of the game before release and during subsequent updates~\cite{DBLP:journals/sbcjis/DuarteMDNE24}. Unlike conventional software applications, game software requires particular attention to gameplay mechanics, user interface, and performance aspects~\cite{DBLP:conf/dsa/ZhangZAI23}.

As the complexity of modern games continues to grow, traditional manual testing methods are increasingly challenged by the vast range of possible game states, actions, and narrative branches. In response, researchers and developers have explored various automated and AI-driven testing approaches. The following sections detail the evolution of these methods, starting with basic scripted automation, then moving on to AI-assisted techniques, and ultimately examining strategies designed to handle complex and dynamic testing scenarios.

\emph{Automation in Game Testing.}  
For relatively simple and fixed game scenarios, automated testing is often carried out by scripting predefined actions~\cite{spronck2006adaptive,yuechen2020regression}. In this approach, testers simulate human player behaviors by designing possible actions and writing control scripts based on them. However, as game environments grow more complex, the cost and inefficiency of writing such custom scripts increase significantly. To enhance the level of automation in game testing, various artificial intelligence (AI) techniques have been applied in multiple areas, including GUI testing, the feasibility testing of game quest scripts, and functional testing~\cite{zarembo2019analysis}.

\emph{AI-Driven Testing Methods.}  
AI-driven approaches aim to reduce human involvement and improve testing efficiency. A common AI testing method is training agents to mimic human behavior to complete specified quests~\cite{zook2019automatic,ariyurek2019automated,ariyurek2020enhancing}. However, these methods depend heavily on the quality of human-generated experience data and are often constrained by issues related to generalization, limiting their applicability in different testing scenarios. To overcome this dependency on human experience data, some studies have proposed driving agents with predefined rewards to accomplish specific behaviors, thus achieving the testing objectives~\cite{loubos_automated_2018,pfau2017automated}. Nevertheless, designing reward functions remains scenario-specific, and these reward-driven tests share similarities with imitation learning. {\color{black}Their effectiveness is limited because outcomes depend heavily on how rewards are designed. This makes it hard to achieve the comprehensive and flexible exploration required for thorough testing.}


\emph{Testing Strategies in Complex Environments.}  
In addition to simple goal-oriented testing approaches, another direction in automated game testing is training agents to explore game environments~\cite{roohi_review_2018}, attempting to detect errors in specific game states. Wuji~\cite{zheng2019wuji} uses evolutionary algorithms to generate diverse agents, making it easier to explore new states and thus increasing the likelihood of discovering errors. Other works~\cite{guerrero-romero_using_2018} explore the environment through agents with different preferences, but these preferences are defined by humans and do not change during the exploration process. Additionally, graph-based~\cite{ammanabrolu_playing_2018,debnath2019review} and tree-based methods~\cite{auer2002finite,silver_mastering_2017} have been applied in game exploration. With the rise of quest descriptions in text format, text classification~\cite{joulin_bag_2017,aggarwal_survey_2012} and other natural language processing methods~\cite{he_deep_2015,chen_learning_2011} can also be leveraged in game testing.
Furthermore, in our previous study collaborated with NetEase~\cite{hu2024language}, traditional language models (e.g., FastText, LTSM) have been explored to accelerate smoke testing by combining quest description texts with online planning. However, unlike LLMs, these traditional models have limited performance and are not suitable for complex quest descriptions and multi-ending quest scenarios.


\subsection{Applications of LLMs in Games}
Due to LLM's exceptional capabilities in natural language understanding, contextual reasoning, and inference, they are considered highly promising for enhancing and streamlining the design and development of games. Existing research has already integrated LLMs into numerous in-game applications, including acting as players~\cite{wang2023voyageropenendedembodiedagent}, serving as NPCs~\cite{warpefelt2017model}, assisting players as assistants~\cite{rist2024using}, providing game commentary~\cite{ranella2023towards}, functioning as game masters in Tabletop RPGs~\cite{you2024dungeons}, analyzing player data~\cite{wang2024player2vec}, designing level difficulty~\cite{todd2023level}, and more recently even operating as game mechanics~\cite{oasis2024} through multimodal models that enable direct player interaction.

\emph{LLMs as Players in Games.}
Among the aforementioned applications of LLMs in games, the use most similar to this research is the use of LLMs as players in games, i.e., allowing LLMs to act as players who can make decisions and perform actions to complete specific quests within a game. A key challenge in enabling LLMs to "play" games lies in converting game states and actions into input formats that LLMs can process efficiently. For instance, GATO~\cite{Gato} leverages a multimodal dataset for training, directly processing visual inputs to achieve near-human performance in Atari games. However, this approach requires extensive, high-quality game video datasets, posing significant challenges in terms of cost and scalability.

As an alternative, some studies propose indirect interaction methods where LLMs do not directly engage with the game environment but instead rely on intermediary layers, such as predefined scripts or interpreters. This setting has been particularly applied in open-world games like Minecraft.
For instance, JARVIS-1~\cite{wang2023jarvis} integrates multimodal models (MLMs) with LLMs to perceive the environment, plan actions, and execute complex quests in an open-world setting via predefined code layers. This enables efficient planning and robust quest execution in the Minecraft environment.
Similarly, VOYAGER~\cite{wang2023voyageropenendedembodiedagent} utilizes GPT-4's code generation capabilities to interact with the Mineflayer API. By dynamically generating code snippets, VOYAGER can complete advanced instructions (e.g., “find the nearest tree”) and continuously expand its skill library to improve quest efficiency.
Plan4MC~\cite{yuan2023skill} introduces a hierarchical multi-quest framework that combines reinforcement learning with LLMs. By decomposing complex open-world quests into manageable sub-skills, Plan4MC significantly improves sample efficiency and quest completion rates. DECKARD~\cite{nottingham2023embodied} takes a similar approach, combining LLMs and RL to hypothesize and verify abstract world models (AWMs), enhancing exploration and sample efficiency in quests such as item crafting within Minecraft.
{\color{black}
In addition to acting as in-game players or assistants, LLMs have also been explored as strategic guides for reinforcement learning agents during the pretraining phase. For example, Du et al.~\cite{du2023guiding} propose a novel framework (ELLM) where LLMs generate semantically meaningful, commonsense-aligned goals for RL agents to pursue. These goals, derived from natural language descriptions of environment states, serve as intrinsic rewards that steer the agent’s exploration behavior toward task-relevant regions of the state space. Compared to traditional intrinsic motivation methods such as novelty-based exploration, this LLM-guided approach encourages more purposeful behaviors that are better aligned with downstream tasks. 
}
{\color{black}

\emph{LLM-based Planning and Reasoning Frameworks.}  
Beyond game-specific applications, recent advances in LLM-based planning include general-purpose reasoning frameworks such as ReAct~\cite{yao2022react}, which interleaves reasoning traces with task-specific actions to improve interpretability and reduce hallucination, and Tree of Thoughts~\cite{yao2023tree}, which explores multiple reasoning paths through deliberate search and backtracking. While these methods demonstrate strong problem-solving capabilities in tasks like question answering and mathematical reasoning, they primarily target reasoning-heavy scenarios and operate through linear or tree-based exploration without explicit mechanisms for multi-ending coverage or state redundancy elimination. Additionally, code-based action frameworks such as CodeAct~\cite{wang2024codeact} unify action spaces by generating executable Python code, enabling flexible tool composition and dynamic action revision. However, such approaches are designed for general API interactions and lack the domain-specific optimizations required for game testing workflows, such as structured path replay, coverage analysis, and integration with QA pipelines.
}
Compared with the above studies, our research differs in several key aspects. 
First, we approach the problem from a developer's perspective rather than that of a player, focusing on improving testing processes rather than simulating player behavior. 
Second, we incorporate a broader range of inputs beyond direct UI interactions, such as game metadata and development-side information, allowing our framework to leverage richer contextual details for more targeted and efficient testing. 
Third, our framework places a strong emphasis on path diversity and reliability, addressing the unique demands of multi-ending quests and ensuring that smoke testing captures the full complexity of branching narratives and gameplay possibilities.

\section{Problem Formulation}
\label{sec:problem}
This section first introduces the core terminology used in this paper to describe the game environment. Based on these core terms, it then provides a problem definition constructed on the Markov Decision Process (MDP).

\subsection{Terminology and Notations}
\label{sec: problem}
This paper targets the game that has Markovian characteristics, where the interaction between the agent and the game environment can be considered as a variation of MDP. 
To facilitate subsequent discussion, we outline below the terminology, concepts, and notations used to describe the interaction model between the intelligent agent and the game environment.
\begin{itemize}
    \item \( S \) is the set of states, representing all possible states in the game, such as the player's position,various status and surrounding environment.
    \item \( A \) is the set of actions, representing all possible actions the intelligent testing agent can perform, such as moving, picking up items, and interacting with NPCs.
    \item \( T \) is the state transition function, \( T: S \times A \to S \), which defines how the system transitions to a new state \( s_{t+1} \) when the system is in state \( s_t \) and the agent performs action \( a_t \) at time step \( t \).
    \item \( Q \) is the quest descriptions, providing context and objectives for the tasks the intelligent agent needs to complete. These descriptions are often presented in natural language or other structured formats, guiding the player’s decision-making process.
    \item \( E \) is the quest completion state set, representing the states where the quest is considered complete, such as achieving a certain game ending.
    \item \( I \) is the set of auxiliary information related to the game’s development, such as source code, level design, and contextual data. This information is often represented in natural language or code and serves as a resource for the testing agent to guide its interactions within the environment.
\end{itemize}

At each time step \( t \), the intelligent testing agent selects an action \( a_t \in A \) based on the current state \( s_t \in S \) and quest description \( Q \). Upon performing the action, the system transitions to a new state \( s_{t+1} = T(s_t, a_t) \). This process repeats iteratively until the agent reaches a quest completion state \( s_t \in E \). The primary goal of the agent is to explore all possible state-action paths leading to a state within the quest completion set \( E \), ensuring comprehensive exploration and validation of the quest execution. Formally, this quest verification process can be framed as a state-action path search problem, where the agent begins at the initial state \( s_0 \in S \), selects appropriate actions based on the quest description, and transitions through successive states until a quest completion state \( s_t \in E \) is reached.

\subsection{Problem Definition}
While the MDP formulation provides a fundamental model of states and actions, complex quests often require a structured approach to bridge from abstract objectives to concrete executable actions. Drawing inspiration from Hierarchical Task Network (HTN) planning, we introduce a hierarchical refinement process, which not only refines the quest description \( Q \) step-by-step but also leverages the auxiliary information \( I \) to guide each layer of refinement. This ensures that the derived tasks and actions remain logically consistent with both the quest goals and the underlying game design.

\begin{definition}[Game-Agent Interaction]
The game environment is defined as a 6-tuple $M=\langle S, A, T, Q, E, I \rangle$, where $S$ is a set of states, $A$ is a set of actions, $T$ is a state transition function, $Q$ is descriptions, $E$ is a quest completion state set, and $I$ is a set of auxiliary information related to the game.
The intelligent agent interacts with the game environment based on a hierarchical design. 
The hierarchical design consists of three levels from top to bottom: \high{} $\hat{\hat{\pi}}$, \low{} $\hat{\pi}$, and action-level execution $\pi$.
\high{} is defined as $\hat{\hat{\pi}}: Q\times I\rightarrow \mathcal{G}$, where $\mathcal{G}$ is a set of abstract quest goals, and each abstract quest goal $g_{h}\in\mathcal{G}$ can correspond to one or more trajectories from the starting state to the completion state.
\low{} is defined as $\hat{\pi}:\mathcal{G}\rightarrow \mathcal{X}$, where $\mathcal{X}$ is a set of pattern sequences.
A pattern $g_{l}\in\mathcal{G}_{low}$ corresponds to a sequence of actions with clear semantics in the game, $\mathcal{G}_{low}$ is the space of the pattern.
Action-level execution is defined as $\pi: \mathcal{G}_{low}\times S \rightarrow A$.
 $\hfill\square$
\end{definition}




Through this hierarchical refinement, the agent does not need to reason over the entire action space directly. Instead, it begins with high-level quest information, step-by-step refines that information into actionable strategies and tasks, and ultimately arrives at fully specified game actions. This approach both simplifies the planning process and ensures that the testing agent can systematically explore and validate complex, multi-ending quests while maintaining coherence with the underlying game design.
\section{\methodname{}}
\label{sec:proposal}
In this section, we first introduce the overview \methodname, followed by a detailed explanation of each core component of the framework. Finally, we provide a concise description of the design of the LLM prompts.

\subsection{Overview of \methodname{}}
To address the challenges posed by smoke testing multi-path and multi-ending quests in modern games, we propose a hierarchical planning framework, \methodname{}, that employs LLMs for multi-level planning. Specifically, we decompose the quest planning process into four hierarchical layers—root task, method, primitive task, and input action—to ensure efficient and diverse path exploration while maintaining precise task execution. By further integrating the state-action transition graph, we maximize path coverage and effectively eliminate redundant routes.

\begin{figure*}[ht]
    \centering
    \includegraphics[width=0.85\linewidth]{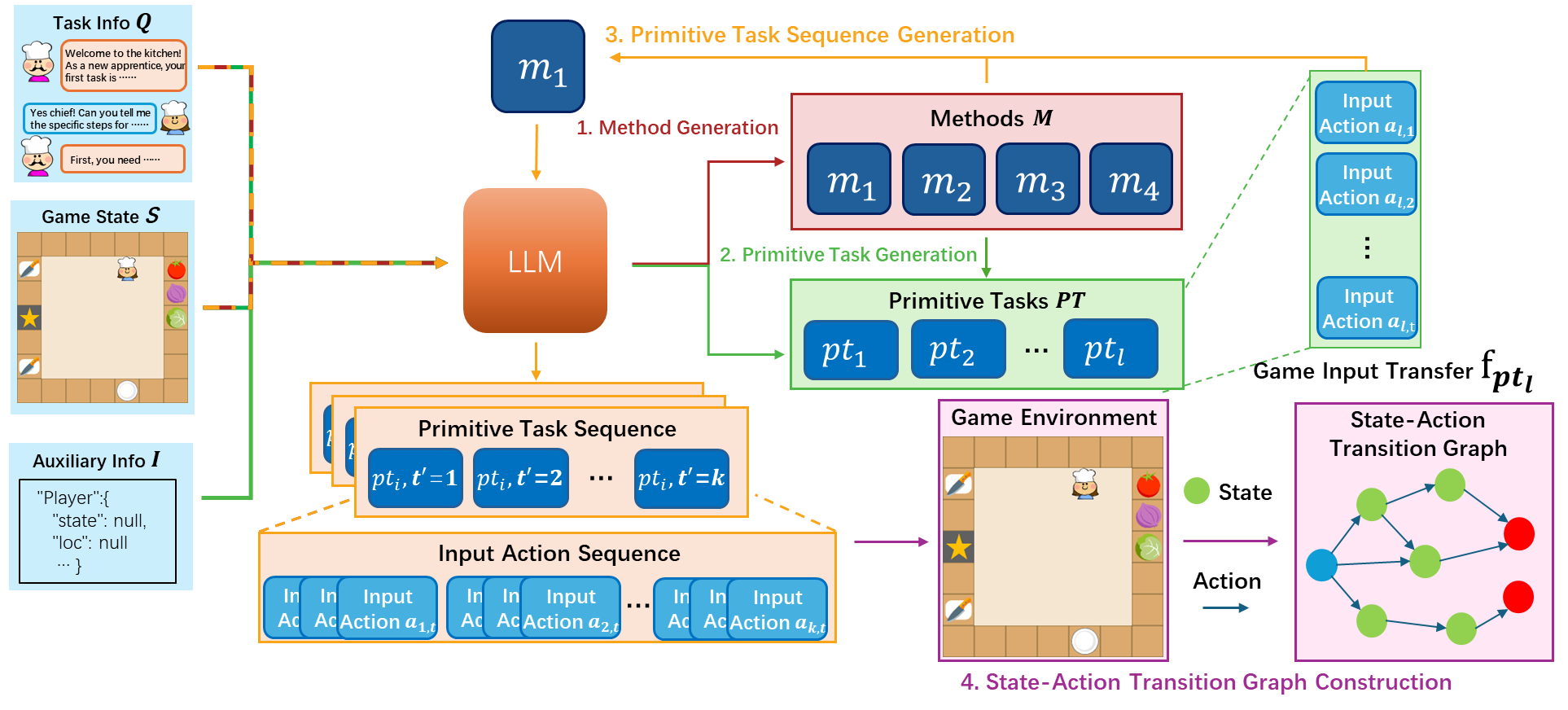}
    \caption{The overview of \methodname{}.}
    \label{fig:framework}
\end{figure*}

The main steps of the framework are illustrated in Fig.~\ref{fig:framework}:
\begin{enumerate}
    \item \textbf{\High{} Generation}:  
    At this stage, the goal is to break down a complex, multi-ending quest into a set of high-level \high{} strategies that serve as abstract solutions for fulfilling the quest.  
    For example, in a kitchen-themed game scenario where the quest is to submit a tomato-containing dish, the framework can derive multiple \high{} from the quest information and game design details—such as “prepare an onion-tomato salad” or “prepare a lettuce-tomato salad.” By doing so, subsequent planning processes can consider various approaches to completing the quest rather than relying on a single fixed pathway.

    \item \textbf{\LOW{} Generation and Input Action Mapping}:  
    Once a \high{} is determined, the next step involves decomposing it into smaller, more reusable \low{}.  
    Continuing the previous example, to achieve the \high{} “prepare an onion-tomato salad,” the system might generate \low{} such as “pick up a tomato” or “chop the onion.” A function generated by the LLM then maps each \low{} to the corresponding game input actions (e.g., pressing a sequence of specific keys), enabling the framework to execute these tasks within the game environment.

    \item \textbf{\LOW{} Sequence Assembly}:  
    In this step, multiple \low{} are combined into consecutive sequences to fulfill the selected \high{} and ultimately achieve the game goal.  
    For instance, to complete the onion-tomato salad, the framework might produce a sequence like: “pick up a tomato,” then “chop the tomato,” next “pick up an onion,” then “chop the onion,” and finally “assemble the ingredients on a plate.” Multiple such sequences may be generated to cover different route of reaching the same goal, ensuring comprehensive test coverage.

    \item \textbf{State-Action Transition Graph Construction}:  
    As each \low{} sequence is executed, the system records the corresponding game states and their transitions, ultimately constructing a graph that captures all explored paths.  
    In the salad-making scenario, each input action required by a \low{} (e.g., the actions needed to pick up an ingredient) leads to a new game state. By tracking these states and their triggering actions, the system builds a state-action transition graph. This facilitates systematic path analysis and metric extraction while eliminating redundant routes, thus supporting targeted testing of critical branching scenarios in multi-ending quests.
\end{enumerate}

\subsection{Step1: \High{} Generation}
In the \methodname{} framework, we treat multi-ending quests as compound tasks, which are inherently complex goals composed of multiple interlinked subgoals and branching solution paths. To effectively address these complexities, we first introduce \high{}s as high-level strategies guiding the pursuit of the overarching quest goal.

The first stage of our framework focuses on generating \high{}s by analyzing quest descriptions and auxiliary information. Utilizing the natural language understanding and reasoning capabilities of LLMs, we decompose the compound quest into abstract \high{}s. These \high{}s represent diverse pathways to quest completion and provide a foundation for subsequent lower-level planning.

The \high{} generation process can be modeled as mapping quest descriptions and auxiliary information to a set of abstract \high{}s. Let \( Q \) represent the quest descriptions, \( I \) represent auxiliary information (e.g., game design information), and \( P \) represent the prompt designed to guide the LLM in generating objectives. Then, the \high{} generation process can be expressed as:

\[
(Q, I, P) \xrightarrow{\text{LLM}} M = \{m_h \mid h=1,\dots,n\}
\]

where \( M = \{m_h\} \) is a set of \high{}s generated by the LLM. Each \( m_h \in M \) represents a possible method to complete the quest. Notably, \( m_h \) is generated by the LLM based on its understanding of the quest descriptions and auxiliary information and primarily serves as guiding reference information for the subsequent planning phase, input back into the LLM. \( \text{LLM} \) utilizing a large language model to derive \( M \) from the inputs \( Q \), \( I \), and \( P \). \( P \) is the prompt serving as a structured input designed to guide the LLM in generating objectives. It provides contextual information and instructions that help the LLM align its output with the requirements of the game environment and the specific objectives to be achieved.


This abstraction simplifies the solution space by reducing the problem's dimensionality while retaining the flexibility required for high-freedom quests. The \high{}s act as guiding nodes in a broader quest graph, structuring exploration and ensuring systematic coverage of possible paths.

\subsection{Step2: \LOW{} Generation and Input Mapping}
Based on \high{}s, the framework uses LLMs to generate general \low{}s that serve as the fundamental units for planning and are mapped into actionable game input actions.

Directly using LLMs to generate every specific game action is inefficient. In grid-based RPG games, for instance, many player inputs involve repetitive actions, such as moving between locations, which do not require advanced reasoning. Generating such actions with LLMs increases computational costs without significant benefits. Furthermore, as LLMs rely on text-based inputs and outputs, they are better suited for generating discrete inputs, limiting their applicability to games requiring continuous inputs.

To address these challenges, the framework generates \low{}s instead of direct game input actions. By combining with game development information, \low{}s are efficiently mapped to game input actions. This method reduces computational overhead and expands the framework’s applicability to both discrete and continuous input games.

\subsubsection{\LOW{} Generation}  
\Low{}s are designed for composability, enabling \high{}s to be achieved by combining and arranging these general-purpose building blocks. Each \low{} is defined as a structured element to ensure clarity, reusability, and compatibility with subsequent \low{} sequence planning and input mapping processes.  
Formally, the \low{} generation process can be expressed as:
\[
(Q, I, M, P) \xrightarrow{\text{LLM}} PT = \{pt_l \mid l=1,\dots,m\}
\]
where \( PT = \{pt_l\} \) is the set of \low{}s.
Each \( pt_l \in PT \) is a general-purpose task designed to be composable and flexible. These \low{}s act as reusable building blocks that can be combined to achieve various \high{}s. \( I \) is auxiliary information that includes game design introductions and other contextual information related to the instruction.

Each \low{} \( pt_l \in PT \) is defined as a triplet:
\[
pt_l = \langle \text{Desc}_l, \text{Pre}_l, \text{Effect}_l \rangle
\]
where \( \text{Desc}_l \) is a textual description of the task, typically consisting of a verb and a target noun (e.g., “put the dish on the table”), ensuring the basic and general nature of the tasks. \( \text{Pre}_l \) represents the preconditions that must be satisfied before execution to ensure the validity of the task (e.g., “player must have a dish”), and \( \text{Effect}_l \) denotes the effects after completing the \low{}, providing a reference for subsequent steps (e.g., “dish on the table”). 

This structured representation ensures that each \( pt_l \) is clear, reusable, and compatible with subsequent steps. Moreover, while \( \text{Pre}_l \) and \( \text{Effect}_l \) provide important guidance for planning, they act as soft constraints rather than strict requirements, ensuring flexibility in action planning and execution.

\subsubsection{Input Mapping}
Once \low{}s are generated, they are mapped to actionable game input actions through a mapping function \( f_{pt_l} \). This mapping function is generated by the LLM based on the \low{} \( pt_l \) and auxiliary game development information \( I \). Formally, the mapping function can be expressed as:
\[
(pt_l, I, P) \xrightarrow{\text{LLM}} f_{pt_l}
\]
For each \( pt_l \in PT \), the mapping function \( f_{pt_l} \) translates the current game state \( s \) into a sequence of game input actions:
\[
f_{pt_l}: s \to \mathcal{A}_l = [a_1, a_2, \dots, a_k]
\]
where \( s \) is the current game state. \( \mathcal{A}_l \) is the sequence of game input actions corresponding to \( pt_l \).

For example, the \low{} “move to target location” might utilize pathfinding algorithms, such as A* or Dijkstra, to compute the optimal route from the player’s current position to the target location. This route is then decomposed into specific game input actions (e.g., “move forward,” “turn right”) and combined with additional commands (e.g., “pick up,” “use”) to produce complete action scripts.

This approach enables the framework to handle a variety of game input action types while ensuring logical consistency and execution efficiency.

\subsection{Step3: \LOW{} Sequence Generation}
After generating the \low{}s, the system utilizes LLMs to generate detailed task sequences for each \high{}. These sequences serve as actionable plans to accomplish the specified \high{}s.

The sequence generation process can be formalized as:
\[
(s_t, Q, m_h, PT, P) \xrightarrow{\text{LLM}} \text{Sec}_{m_h}
\]
where \( s_t \) is the current game state, \( \text{Sec}_{m_h} \) is the \low{} sequence generated for the \high{} \( m_h \).
The sequence consists of a set of \low{}s:
\[
\text{Sec}_{m_h} = \{pt_{l,1}, pt_{l,2}, \dots, pt_{l,k}\}
\]
Here, \( pt_{l,t'} \) represents the \(t'\)-th \low{} chosen to accomplish the \high{} \( m_h \).
    
To ensure the accuracy of the generated action sequences, we incorporate Chain-of-Thought (CoT)~\cite{NEURIPS2022_9d560961} reasoning and a Reflection Mechanism~\cite{NEURIPS2023_1b44b878} as prompt engineering techniques to optimize the LLM’s output.

CoT guides the model to reason step-by-step, helping it establish logical relationships between actions and avoid reasoning jumps. Specifically, when generating each \( pt_{l,t'} \), the LLM is prompted to simultaneously produce three elements: first, a \low{} \( pt_{l,t'} \), which is a specific objective description indicating the task to be achieved (e.g., “move to the kitchen”); second, a state analysis \( s'_{t'} \), which details how this action affects the current game state and any resulting changes (e.g., “player reaches the kitchen”); and finally, a next-step plan \( P_{t'+1} \), which outlines the subsequent action or strategy based on the updated state and the ongoing objective (e.g., “open the refrigerator and take out the lettuce”).

Through CoT, the LLM progressively derives various stages of the quest, ensuring that each task aligns with the objective. This structure allows the LLM to consider the current state while anticipating the consequences of each action, improving the accuracy of sequential action planning.

Additionally, to further enhance the accuracy and effectiveness of the \low{} sequences, we introduce a Reflection Mechanism to refine the task plan. This mechanism follows three sequential steps. First, the LLM generates an initial sequence of \low{}s based on the given task description and the current state. Second, it undergoes a reflection phase, during which the initial sequence is carefully analyzed to identify and pinpoint any logical inconsistencies, inefficiencies, or errors. Finally, in the revision phase, the LLM revises and optimizes the initial sequence based on the issues identified during reflection, ensuring that the resulting task plan is more accurate, coherent, and efficient.

By combining CoT reasoning with the Reflection Mechanism, the system ensures the accuracy and coherence of action planning. This approach not only improves the quality of the initial objective sequences but also refines them iteratively, enhancing task completion efficiency and success rates. Ultimately, this mechanism ensures that the execution order and state transitions align with expectations, maximizing the accuracy of quest execution in complex scenarios.

For each \high{}, multiple \low{} sequences are generated to enrich the state-action transition graph construction, and by leveraging the mapping function \( f_{pt_l} \) associated with each \( pt_{l,t'} \) in the sequence \( \text{Sec}_{m_h} \), these sequences are translated into corresponding game action inputs for interaction with the game environment.

\subsection{Step4: State-Action Transition Graph Construction}

As the system executes action sequences, it dynamically constructs a state-action transition graph that captures all explored game states and their transitions. In this graph, each node represents a unique game state \(s_i\), and is encoded as a fixed-length vector \( \mathbf{s_i} \in \mathbb{R}^d \) to ensure distinct representation. The state vector \( \mathbf{s_i} \) encapsulates all the relevant information about the state at any given point in the game, ensuring that each state is uniquely identifiable.

Edges in the graph represent the actions \(a_i\) that transition between game states. Formally, a state transition can be described as:

\[
s_{i} \xrightarrow{a_i} s_{i+1}
\]


The state-action transition graph serves several potential purposes in the framework:
\begin{itemize}
    \item Path Analysis: The graph allows the system to identify all feasible paths from the initial state \(s_0\) to the quest completion state \(s_n \in E\). Specifically, the system can trace different sequences of actions \( \{a_1, a_2, \dots, a_n\} \) that lead to a final objective state \(s_n \in E\). The total number of such paths can be computed by traversing the graph and counting all unique sequences of edges from \(s_0\) to \(s_n \in E\).
    
    \item Redundancy Elimination: As the system explores the game space, redundant or invalid paths that do not lead to quest completion or that revisit previously explored states can be filtered out. This ensures more efficient exploration by removing circular paths or states that do not contribute to objective achievement.

    \item Multi-Ending Coverage: For quests with multiple possible outcomes, such as reaching different destinations in a game, the state-action transition graph highlights distinct paths leading to each possible endpoint. For instance, a player might be tasked with traveling to either a forest or a mountain, with each destination requiring different actions, such as selecting specific routes or gathering necessary resources along the way. The graph captures these variations by mapping out the unique steps required for each destination, dynamically adapting to the quest's requirements. By reflecting different choices and their resulting paths, the graph ensures a comprehensive exploration of all possible outcomes, enabling thorough testing and validation of multi-ending scenarios.

    \item Enhanced Interpretability: Beyond its abstract vector representations, the state-action transition graph can be integrated with visualization tools to support reverse mapping from state vectors back to their corresponding in-game snapshots. By complementing each node with visual cues (e.g., game screen captures) or descriptive state information, developers can more intuitively identify and interpret each state and its transitions. {\color{black}This visual linkage helps developers better understand gameplay, identify issues, and communicate results, making debugging more efficient.}
\end{itemize}


\subsection{Prompt Design}
\begin{figure*}[ht]
    \centering
    \includegraphics[width=0.80\linewidth]{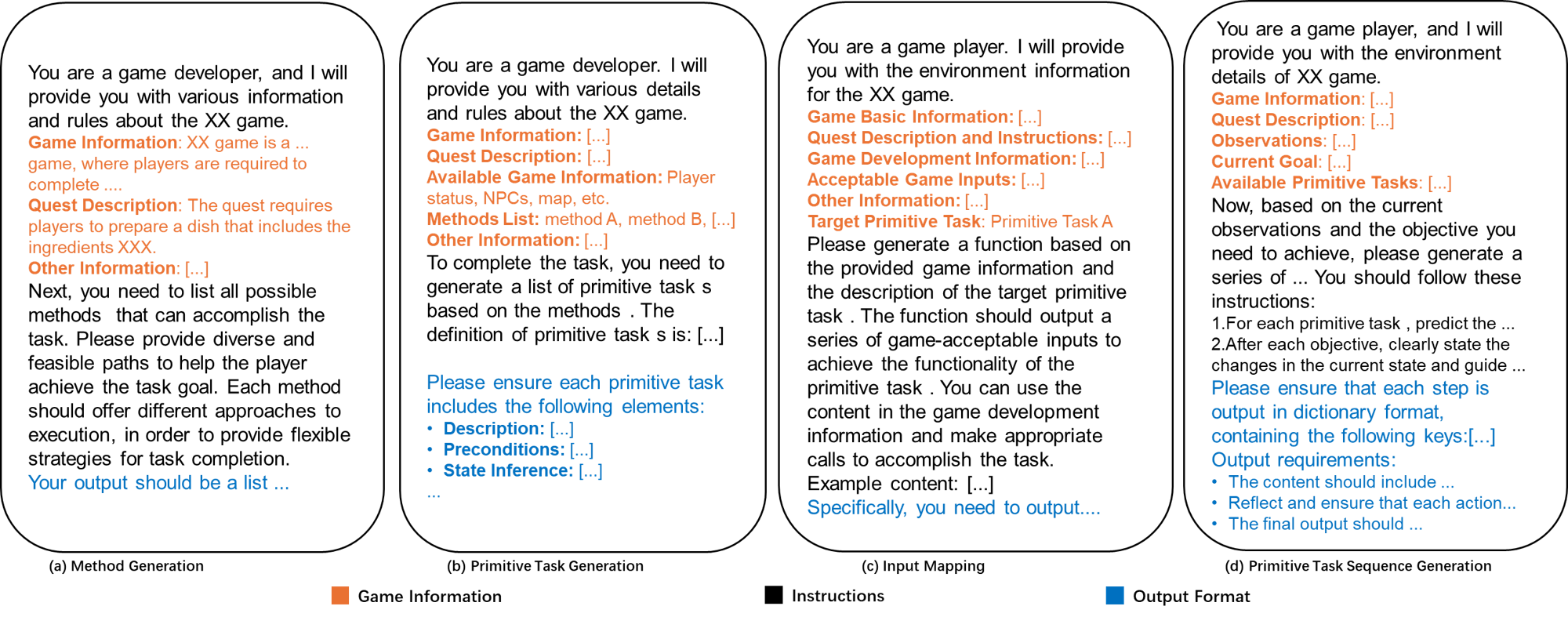}
    \caption{Prompt example for the \methodname{}}
    \label{fig:prompt}
\end{figure*}
Figure~\ref{fig:prompt} shows examples of prompts used in this study. 
We systematically design the LLM prompts by considering the following key elements.

\textbf{Game Information and Quest Background}.
The primary task of the prompts is to ensure that the LLM receives sufficient game information and quest context to correctly interpret the quest and make appropriate decisions. Game information includes the game type, quest objectives, available resources, and player states, all of which help the LLM quickly understand the game environment before starting the quest. When designing the prompts, we distinguish between two perspectives: the developer's perspective and the player's perspective. The developer's perspective focuses on the technical details of the game, such as quest structure, level design, and possible action sequences, while the player's perspective focuses on the player's actual interactions with the game environment, assisting the LLM in generating action objectives and input mappings that align with the player's needs.
This distinction in perspectives ensures that the LLM can generate contextually accurate outputs at different stages of the quest. In the action objective generation phase, the developer's perspective helps the LLM generate valid and reliable objectives to support subsequent testing. In the action sequence generation phase, the player's perspective provides necessary contextual information, helping the LLM accurately plan the sequence of actions and simulate player behavior. 

\textbf{Prompt Instructions}. The instructions within the prompt are key to guiding the LLM in generating the action sequences. The instructions vary according to the objectives of different stages in the quest. To ensure the LLM accurately understands the quest and generates high-quality steps, the prompts must possess several key characteristics: first, they should be clear and concise, avoiding unnecessary complexity to reduce comprehension errors; second, tasks should be broken down into stages to allow the LLM to work step-by-step and avoid information overload; third, the prompts must be closely aligned with the game environment, ensuring that the actions generated by the LLM are consistent with the actual game state, resources, and objectives.
Furthermore, the introduction of CoT and Reflection mechanisms can significantly improve the quality of quest generation. CoT helps the LLM verify the necessity of each action through step-by-step reasoning, reducing potential errors in quest planning. The reflection mechanism allows the LLM to identify and correct logical flaws during the generation process, improving the overall accuracy and coherence of the generated sequence.

\textbf{Output Format Control}.
To ensure that the LLM’s output is structured, analyzable, and efficient for subsequent processing, we impose strict output format requirements in the prompts. Each stage of the quest output should follow a consistent format, particularly in the generation of \low{}s and action sequences, where the output is presented in a dictionary structure. The dictionary must contain key information such as ``Action Description,'' ``State Analysis,'' and ``Next Step Planning,'' clearly indicating the purpose, preconditions, and expected outcomes of each operation for subsequent analysis and execution.
This format control not only ensures the logical coherence and traceability of quest execution but also enables effective quest analysis and testing through structured output data. 


\section{Evaluation}
\label{sec:evaluation}
{\color{black}
This evaluation is designed to address the following research questions (RQ):
\begin{itemize}
    \item RQ1: How consistently can \methodname{} complete quests, and how concise are its successful action plans, compared to baseline methods?

    \item RQ2: How effectively can \methodname{} discover and cover all designated quest endings?

    \item RQ3: How does \methodname{} compare with baseline methods in terms of the total number of environment interaction steps required during the testing process?
\end{itemize}

\subsection{Experimental Setting}
\label{sec:exp_setup}

\textit{Evaluation Metrics}.
We adopt the following quantitative metrics for the above research questions.
For RQ1, we adopt two key metrics. The first is the completion rate, which measures the proportion of test runs in which the quest is successfully completed. The second is the average number of steps required to complete the quest, calculated only over successful runs, which reflects the length of the action path needed to achieve completion.
For RQ2, we count the number of times each possible ending (Ending A, B, or C, with increasing difficulty) is achieved among all successful runs, and compute their corresponding proportions. 
For RQ3, we evaluate the total number of environment interaction steps incurred throughout the entire testing process, including both training and testing phases.

\textit{Baseline Methods}.
We select four representative baselines. All methods operate on the same underlying environment dynamics and action semantics, while using method-appropriate observation representations:

{\color{black}
(i) Random Algorithm~\cite{hu2024language,li2022gbgallery}: Randomly samples actions from the available action space until the quest is completed or the maximum step limit is reached, serving as a difficulty benchmark for uninformed exploration. The maximum step limit is 5{,}000 steps for Mini RPG, 10{,}000 steps for Overcooked Plus, 2{,}000 steps for Minecraft, and 200 steps for PySC2. Within one iteration, each quest/task is sampled 30 times.

(ii) Genetic Algorithm (GA)~\cite{holland-genetic-algorithms-1992}: A classic search-based method that evolves fixed-length action sequences to optimize quest completion. The sequence length is 100 for Mini RPG, 1{,}000 for Overcooked Plus, 50 for Minecraft, and 200 for PySC2. The population size is 100; the maximum number of generations is 150 for Mini RPG/Overcooked Plus/Minecraft and 50 for PySC2; the crossover operator is Xovdp with rate 0.7; the mutation operator is Mutbga with rate 0.01, shrink parameter 0.5, and gradient parameter 20. An early stop mechanism terminates evolution when no fitness improvement is observed for 10 consecutive generations. Within one iteration, each quest/task is trained 10 independent times, and the top 3 individuals in the final population are kept for evaluation.

(iii) Proximal Policy Optimization (PPO)~\cite{PPO}: A widely used policy-gradient reinforcement learning algorithm that optimizes a policy network to maximize expected return. We use a multi-layer perceptron with vectorized game states as input and design environment-specific reward functions. Training uses a maximum budget of 500{,}000 timesteps, with early stopping if no improvement is observed for five consecutive evaluations. Hyperparameters are: learning rate $3 \times 10^{-3}$, rollout steps $n = 2048$, batch size 64, discount factor $\gamma = 0.99$, GAE parameter $\lambda = 0.95$, clip range 0.1, value function coefficient 0.5, entropy coefficient 0.01, optimization epochs 10, and gradient clipping threshold 0.5. Within one iteration, each quest/task is trained with 10 independently initialized models, and each trained model is evaluated with 3 rollout runs. The reward settings are as follows: Mini RPG: step penalty $-0.1$, subquest completion reward $+100$; Overcooked Plus: step penalty $-0.1$, ingredient processing reward $+10$, quest completion reward $+1000$, incorrect submission penalty $-20$; Minecraft: step penalty $-0.1$, obtaining required item reward $+10$, quest completion reward $+1000$; PySC2: a hybrid reward with terminal outcome reward plus dense shaping feedback based on mission progress, reduction of enemy key targets, battlefield advancement, and friendly-unit preservation, following prior PySC2 reward design practices~\cite{LLM-PySC2}. Inference uses stochastic action sampling to encourage exploration.

(iv) ELLM-PPO~\cite{du2023guiding}: An enhanced PPO method that introduces LLM-assisted exploration. GPT-4o generates natural language subgoals based on the current state description and quest specification. An intrinsic reward is computed as the semantic similarity between state transition descriptions and generated subgoals. The agent is pretrained with this augmented reward and then fine-tuned with standard PPO. All RL training hyperparameters (learning rate, batch size, network architecture, etc.) are kept identical to the PPO baseline for fairness, with the only difference being the addition of LLM-generated intrinsic rewards during the pretraining phase. The same 10-model, 3-rollout protocol is used within one iteration.

(v) \methodname{} (Ours): Uses GPT-4o as the LLM planner, combined with a multi-level planning structure and a state–action transition graph to solve quests. For each \high{}, 30 action sequences are generated and recorded in the state-action transition graph. The graph is then processed by merging identical nodes and removing redundant paths, after which 10 valid quest completion trajectories are randomly sampled as the final output.

All experiments use random seed 42. To estimate statistical uncertainty, the full experiment is repeated 30 times, and all reported values are presented as mean $\pm$ 95\% confidence interval, constructed using the t-distribution. These 30 repetitions are used only for confidence-interval estimation and are not accumulated into Total Steps.
}

\begin{figure}[t]
    \centering
    \subfloat[Mini RPG\label{fig:mini_rpg}]{
        \includegraphics[width=0.45\linewidth]{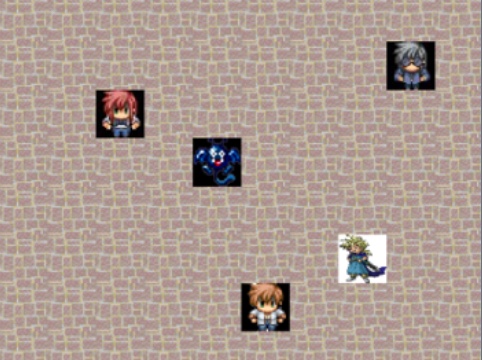}
    }
    \hfill
    \subfloat[Overcooked Plus\label{fig:overcooked}]{
        \includegraphics[width=0.45\linewidth]{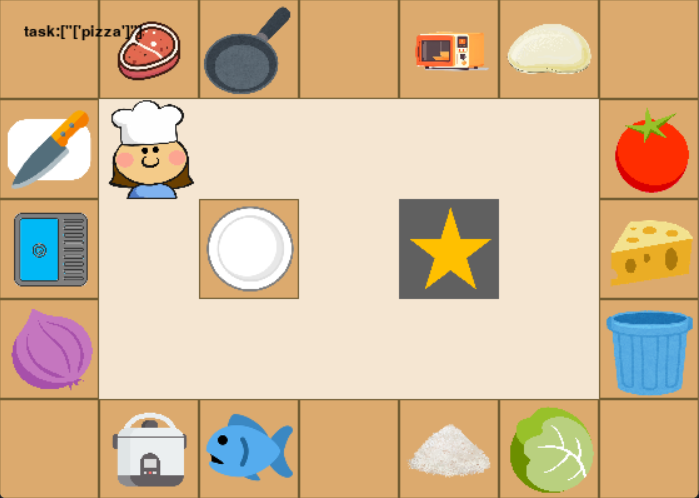}
    }

    \vspace{0.5em}

    \subfloat[Minecraft\label{fig:mc}]{
        \includegraphics[width=0.45\linewidth]{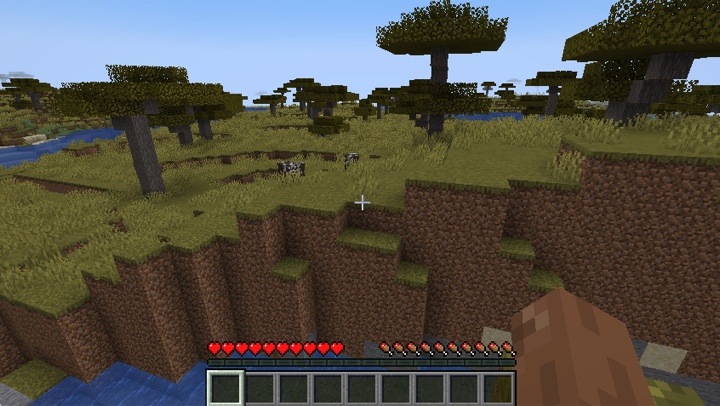}
    }
    \hfill
    \subfloat[PySC2\label{fig:pysc2}]{
        \includegraphics[width=0.45\linewidth]{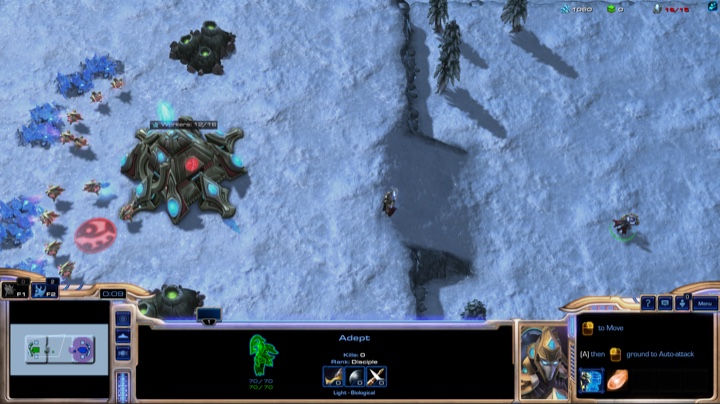}
    }

    \caption{Representative screenshots of the four evaluation environments.}
    \label{fig:envs}
\end{figure}

To comprehensively evaluate our method, we conduct experiments across {\color{black}four} distinct game environments: {\color{black}Mini RPG, Overcooked Plus, Minecraft, and a PySC2-based RTS scenario}. These environments are chosen to cover a diverse range of game genres, quest structures, and difficulty characteristics.

\textit{Mini RPG Game}.
Mini RPG is a simplified, linear, story-driven role-playing game~\cite{hu2024language}, designed to evaluate quest understanding and planning in natural language–rich scenarios. It preserves the core logic of traditional RPGs while simplifying the map and interactions to support rapid quest generation and reproducibility.
The environment consists of a fixed grid map containing NPCs, monsters, the player character, and interactive terrain. Players can interact with NPCs, collect or submit items, engage in combat, and trigger scripted events. The game state is fully observable and represented as a structured dictionary that includes the map layout, character attributes, NPCs, and active quests.
The environment features 9 quests, each with multiple completion routes of varying difficulty. Each route consists of a sequence of subquests described in natural language, requiring the player to follow different action sequences.
Mini RPG Game, as a simple and basic environment, is designed to provide an entry-level testing scenario for the method, verifying whether it can stably achieve objectives in low-complexity quests and establishing a performance baseline for subsequent evaluation in more complex environments.

\textit{Overcooked Plus}.
Overcooked Plus~\cite{Cai2024OvercookedPlus} is an open-source and difficulty-enhanced version of the commercial cooking game Overcooked~\cite{overcooked_team17}, designed to evaluate planning under operational constraints and execution sensitivity.
Players prepare dishes by processing and combining ingredients through actions like chopping, cooking, plating, and serving. The environment enforces strict rules—for example, each station holds only one item, and incorrect placements must be explicitly discarded. It is fully observable, with structured state representations that track the location and status of all relevant objects.
The quest set includes 7 cooking quests, each allowing multiple valid completion strategies reflecting increasing difficulty levels. Harder completions typically require more complex combinations and precise action sequencing.
This environment emphasizes sequential planning under strict rules, requiring the agent to maintain high execution accuracy while rapidly responding to state changes and correcting errors.

\textit{Minecraft}.
Minecraft~\cite{minecraft2009} is a large-scale sandbox environment widely used in AI research due to its open-ended structure and rich combinatorial interaction dynamics.
To support consistent evaluation, we adopt a high-level action abstraction within the \methodname{} framework, where all methods issue discrete, parameterized commands such as ``mine,'' ``craft,'' or ``move.'' These are internally mapped to low-level atomic operations. The environment provides full observability of the agent's position, inventory state, resource locations, and tool status. To ensure comparability across methods, we discretize the action space into a fixed set of parameterized high-level commands.
We define 11 quests inspired by Minecraft’s achievement system, spanning resource gathering, item crafting, and combat. Each quest supports multiple valid solutions of varying complexity, encouraging long-horizon planning and adaptive strategy.
This environment is characterized by difficulties stemming primarily from long dependency chains and a vast state space.

\textit{{\color{black}PySC2}}.
{\color{black}We further introduce a PySC2-based evaluation scenario~\cite{LLM-PySC2,PySC2} to assess scalability in a more complex RTS setting. The scenario contains five tasks under a fixed Protoss-versus-Zerg configuration. Compared with the existing environments, this setting introduces higher decision complexity and requires agents to continuously make real-time decisions based on partially observable in-game information, thereby imposing stronger demands on the online decision-making capability of the methods.}

{\color{black}For observation and action protocols, PySC2 observations are limited to currently visible game-window information, including progress signals, friendly-unit states, visible enemy information, and spatial-position features. PPO and ELLM-PPO consume normalized vectorized observations, while \methodname{} receives semantically structured natural-language observations. The PySC2 action handling follows the same unified high-level action abstraction strategy used in Minecraft to maintain direct comparability across methods. In practice, representative actions include no-op, move-to-position, attack-target, and skill-cast commands, with parameters instantiated as screen/minimap coordinates, target-unit IDs, and lightweight control flags.}

{\color{black}All five tasks are combat scenarios with distinct tactical settings. For each task, we manually define three endings (Ending A/B/C) with increasing difficulty. The ending design is task-specific, and the corresponding achievement conditions vary across tasks, including combat victory (destroying designated enemy units/buildings), defensive victory (protecting designated friendly units/buildings for a required duration), economic victory (reaching target resource thresholds), and draw under predefined special conditions. We note that such explicit multi-ending formulations are not part of the original PySC2 task design. However, because open and experimentally accessible benchmarks that simultaneously offer diverse gameplay settings and explicit multi-ending structures are extremely limited, we construct these endings within the available tasks for controlled evaluation. Although this adaptation does not exactly mirror the original game design, analogous branch-like outcome structures are common in practical game  design.}

{\color{black}The five tasks differ systematically in enemy composition, available friendly-unit combinations, skill configurations, map-space structure, and time pressure. These variations create diverse tactical situations that require agents to continuously interpret partially observable states and make accurate step-level decisions under real-time constraints.}

\subsection{Experiment Results}
\label{sec:exp_results}

\begin{table*}[h] 
\setlength{\tabcolsep}{4pt} 
\centering 
\small 
\color{black}
\caption{Performance Metrics for Quest Completion and Ending Diversity across {\color{black}Four} Game Environments} 
\label{tab:all_game_results_with_ci} 
\begin{tabular}{llcccccc} 
\toprule 
Game & Method & Comp. Rate(\%) & Avg Steps & Ending A(\%) & Ending B(\%) & Ending C(\%) & Total Steps \\  
\midrule 
\multirow{5}{*}{Mini RPG}  
& Random & 25.3 $\pm$ 3.8 & 1463.9 $\pm$ 155.2 & 66.7 $\pm$ 5.1 & 24.4 $\pm$ 4.5 & 8.9 $\pm$ 2.3 & 1,001,046 \\  
& GA  & 54.4 $\pm$ 4.5 & 81.0 $\pm$ 7.8  & 58.5 $\pm$ 5.2 & 24.5 $\pm$ 4.1 & 17.0 $\pm$ 3.3 & 124,086,117 \\  
& PPO & 62.5 $\pm$ 4.8 & 47.2 $\pm$ 3.5  & 74.2 $\pm$ 4.1 & 18.4 $\pm$ 3.2 & 7.4 $\pm$ 2.6  & 10,838,016 \\ 
& ELLM-PPO & 74.6 $\pm$ 3.9 & 53.2 $\pm$ 4.9 & 65.0 $\pm$ 4.4 & 22.1 $\pm$ 3.6 & 13.0 $\pm$ 2.9 & 9,920,000 \\ 
& \methodname{} (ours) & 100.0 $\pm$ 0.0 & 28.3 $\pm$ 1.6 & 40.6 $\pm$ 3.3 & 30.7 $\pm$ 3.1 & 28.7 $\pm$ 2.8 & 6,750 \\  
\midrule 
\multirow{5}{*}{\begin{tabular}[c]{@{}l@{}}Overcooked\\Plus\end{tabular}}
& Random & 49.2 $\pm$ 5.1 & 5686.2 $\pm$ 481.6 & 70.6 $\pm$ 5.8 & 29.4 $\pm$ 5.8 & 0.0 $\pm$ 0.0 & 688,432 \\  
& GA  & 68.5 $\pm$ 4.6 & 753.1 $\pm$ 68.3  & 47.1 $\pm$ 5.5 & 42.9 $\pm$ 5.6 & 10.0 $\pm$ 3.1 & 17,854,231 \\  
& PPO & 56.2 $\pm$ 5.3 & 43.8 $\pm$ 4.1  & 88.8 $\pm$ 2.8 & 9.3 $\pm$ 2.3  & 1.9 $\pm$ 0.9  & 2,905,484 \\   
& ELLM-PPO & 78.4 $\pm$ 4.1 & 79.3 $\pm$ 7.2 & 56.5 $\pm$ 5.1 & 33.6 $\pm$ 4.4 & 9.9 $\pm$ 2.8 & 3,510,323 \\ 
& \methodname{} (ours) & 100.0 $\pm$ 0.0 & 66.0 $\pm$ 3.5 & 23.2 $\pm$ 3.6 & 46.4 $\pm$ 4.3 & 30.4 $\pm$ 3.0 & 15,036 \\  
\midrule 
\multirow{5}{*}{Minecraft}  
& Random & 32.1 $\pm$ 4.4 & 462.4 $\pm$ 91.5 & 74.7 $\pm$ 6.3 & 25.3 $\pm$ 6.3 & 0.0 $\pm$ 0.0 & 431,456 \\ 
& GA  & 55.6 $\pm$ 5.2 & 15.3 $\pm$ 2.4  & 62.5 $\pm$ 6.1 & 25.0 $\pm$ 4.8 & 12.5 $\pm$ 3.3 & 5,262,322 \\ 
& PPO & 43.7 $\pm$ 5.5 & 6.9 $\pm$ 0.9   & 89.9 $\pm$ 3.1 & 7.4 $\pm$ 2.4  & 2.7 $\pm$ 1.2  & 335,163 \\  
& ELLM-PPO & 68.8 $\pm$ 4.4 & 8.6 $\pm$ 1.2 & 69.4 $\pm$ 5.5 & 19.0 $\pm$ 4.1 & 11.6 $\pm$ 3.2 & 293,112 \\ 
& \methodname{} (ours) & 100.0 $\pm$ 0.0 & 7.3 $\pm$ 0.6 & 37.0 $\pm$ 4.5 & 34.3 $\pm$ 4.2 & 28.7 $\pm$ 3.8 & 1,413 \\  
\midrule
\multirow{5}{*}{{\color{black}PySC2}}
& {\color{black}Random} & {\color{black}38.9 $\pm$ 4.0} & {\color{black}89.6 $\pm$ 8.5} & {\color{black}57.8 $\pm$ 5.1} & {\color{black}38.0 $\pm$ 3.9} & {\color{black}4.2 $\pm$ 1.5} & {\color{black}18,320} \\
& {\color{black}GA} & {\color{black}47.3 $\pm$ 4.2} & {\color{black}66.2 $\pm$ 5.7} & {\color{black}56.2 $\pm$ 4.7} & {\color{black}38.1 $\pm$ 3.8} & {\color{black}5.7 $\pm$ 1.8} & {\color{black}12,186,540} \\
& {\color{black}PPO} & {\color{black}64.6 $\pm$ 4.9} & {\color{black}81.4 $\pm$ 6.2} & {\color{black}52.6 $\pm$ 4.6} & {\color{black}40.1 $\pm$ 3.8} & {\color{black}7.3 $\pm$ 2.0} & {\color{black}2,356,904} \\
& {\color{black}ELLM-PPO} & {\color{black}78.6 $\pm$ 3.9} & {\color{black}54.8 $\pm$ 4.5} & {\color{black}46.8 $\pm$ 4.2} & {\color{black}42.9 $\pm$ 3.6} & {\color{black}10.3 $\pm$ 2.1} & {\color{black}2,401,772} \\
& {\color{black}\methodname{} (ours)} & {\color{black}81.2 $\pm$ 3.5} & {\color{black}34.1 $\pm$ 2.0} & {\color{black}48.7 $\pm$ 4.1} & {\color{black}44.9 $\pm$ 3.3} & {\color{black}6.4 $\pm$ 2.0} & {\color{black}25,406} \\
\bottomrule 
\end{tabular} \\
\vspace{2mm}
\footnotesize{\textit{Note:} Ending difficulty is defined by the minimal number of actions required to reach it. Longer sequences indicate higher difficulty.(except PySC2)}
\end{table*}

Table~\ref{tab:all_game_results_with_ci} summarizes the results across {\color{black}all four benchmark environments}. In terms of quest completion rate, \methodname{} remains the strongest method across all environments. In PySC2, it reaches $81.2\%$ completion, outperforming ELLM-PPO ($78.6\%$) and remaining clearly above PPO ($64.6\%$), GA ($47.3\%$), and Random ($38.9\%$). {\color{black}This shows that the semantic planning and graph-based trajectory consolidation in \methodname{} remain effective even when the environment becomes more dynamic and execution-sensitive, although the advantage becomes noticeably smaller in this more reactive setting.}

Regarding average completion steps, \methodname{} also demonstrates a clear advantage. In Mini RPG, it achieves quest completion with an average of only 28.3 steps, significantly lower than all baselines. In Overcooked Plus and Minecraft, although PPO yields lower average step counts (43.8 and 6.9, respectively), a closer examination of the ending distribution reveals a strong bias toward the simplest outcome, Ending A, in both PPO and ELLM-PPO. For example, in Minecraft, PPO achieves Ending A in 89.9\% of successful runs, while the hardest outcome, Ending C, is achieved in only 2.7\%. Since ending difficulty in our evaluation is defined by the required action sequence length, a lower average step count does not necessarily indicate more efficient or effective quest planning. By contrast, \methodname{} achieves a more balanced distribution across all three endings while maintaining relatively low average steps. In PySC2, \methodname{} still attains the shortest successful trajectories (34.1 steps), substantially below ELLM-PPO (54.8), GA (66.2), PPO (81.4), and Random (89.6). This indicates that the framework can still stabilize execution into relatively concise successful plans even in a real-time strategy setting.

In terms of ending coverage, the differences between methods are further amplified in the more challenging environments. While most methods can effectively reach all three endings in the relatively simple Mini RPG game, performance diverges significantly in Overcooked Plus and Minecraft. Random fails to reach any Ending C, and PPO-based methods display a strong bias toward simple paths. GA shows moderate ability to reach high-difficulty outcomes (e.g., 12.5\% for Ending C in Minecraft). In contrast, \methodname{} achieves 30.4\% coverage of Ending C in Overcooked Plus and attains a near-uniform distribution across all endings in Minecraft (37.0\%, 34.3\%, and 28.7\% for Endings A, B, and C, respectively). {\color{black}In PySC2, the pattern becomes more nuanced: \methodname{} still broadens coverage beyond easy endings and achieves the highest proportion on the medium-complexity branch (Ending B: $44.9\%$), but its Ending C proportion ($6.4\%$) remains lower than PPO ($7.3\%$) and ELLM-PPO ($10.3\%$). This suggests that, in the RTS setting, the hardest cases depend less on long dependency-chain reasoning and more on reactive execution under rapidly changing states. Since \methodname{} is not an online planning architecture, its offline long-horizon reasoning is more likely to be affected in this setting, whereas PPO-based methods can occasionally benefit from online adaptation during successful rollouts.}

With respect to testing cost, \methodname{} incurs dramatically fewer total environment interaction steps than the training-based baselines. It requires only 6,750 steps in Mini RPG, 15,036 steps in Overcooked Plus, 1,413 steps in Minecraft, and 25,406 steps in PySC2. By comparison, GA demands over 124 million steps in Mini RPG, PPO consumes over 10 million steps there, and in PySC2 Random, GA, PPO, and ELLM-PPO require 18,320, 12,186,540, 2,356,904, and 2,401,772 steps, respectively. {\color{black}Random is cheaper than \methodname{} in PySC2 because it performs no training, but this lower interaction cost comes with much weaker completion reliability and hard-ending coverage; GA, PPO, and ELLM-PPO all require substantially more interaction than \methodname{}.} Overall, the results show that \methodname{} offers a substantially better performance--cost tradeoff than competitive search and RL baselines, while preserving strong multi-ending exploration capability.
{\color{black}At the same time, interaction efficiency does not fully equal total runtime cost, because \methodname{} also relies on LLM inference. We therefore interpret Total Steps as interaction cost rather than full end-to-end computational cost; a more detailed discussion of inference overhead is provided in the Limitations section.}

\subsection{Ablation Study}
\begin{table*}[h]
 \centering
 \small
 \color{black}
 \caption{Performance comparison of different settings in Overcooked Plus}
 \label{tab:kitchen_ablation}
 \begin{tabular}{llcccccc}
 \toprule
 Setting & Method & Comp. Rate(\%) & Avg Steps & Ending A(\%) & Ending B(\%) & Ending C(\%) & Total Steps \\
 \midrule
 \multirow{5}{*}{Full (Ours)}
  & DeepSeek-V2      & 100.0 $\pm$ 0.0 & 69.2 $\pm$ 4.0 & 20.8 $\pm$ 3.9 & 52.0 $\pm$ 5.2 & 27.2 $\pm$ 4.0 & 16,184 \\
  & Qwen             & 100.0 $\pm$ 0.0 & 77.7 $\pm$ 4.2 & 25.3 $\pm$ 4.1 & 41.8 $\pm$ 5.1 & 32.9 $\pm$ 4.5 & 15,428 \\
  & GPT-4o           & 100.0 $\pm$ 0.0 & 66.0 $\pm$ 4.2 & 23.2 $\pm$ 4.4 & 46.4 $\pm$ 5.5 & 30.4 $\pm$ 4.7 & 15,036 \\
  & GPT-4o-mini      & 100.0 $\pm$ 0.0 & 36.4 $\pm$ 2.1 & 42.0 $\pm$ 10.0 & 58.0 $\pm$ 10.0 & 0.0 $\pm$ 0.0 & 14,951 \\
  & Claude-3.5-Haiku & 100.0 $\pm$ 0.0 & 33.5 $\pm$ 1.4 & 57.0 $\pm$ 10.0 & 43.0 $\pm$ 10.0 & 0.0 $\pm$ 0.0 & 15,363 \\
 \midrule
 \multirow{5}{*}{w/o Graph}
  & DeepSeek-V2      & 26.1 $\pm$ 4.3 & 79.8 $\pm$ 4.3 & 64.3 $\pm$ 9.7 & 20.7 $\pm$ 8.6 & 15.6 $\pm$ 8.5 & 16,149 \\
  & Qwen             & 28.1 $\pm$ 5.6 & 81.2 $\pm$ 2.9 & 45.8 $\pm$ 10.8 & 50.6 $\pm$ 9.1 & 3.6 $\pm$ 3.5 & 15,182 \\
  & GPT-4o           & 46.1 $\pm$ 5.2 & 82.9 $\pm$ 2.6 & 25.7 $\pm$ 9.5 & 59.6 $\pm$ 9.4 & 16.3 $\pm$ 7.6 & 14,823 \\
  & GPT-4o-mini      & 5.6 $\pm$ 2.3  & 69.8 $\pm$ 4.1 & 90.7 $\pm$ 12.2 & 10.6 $\pm$ 13.8 & 0.0 $\pm$ 0.0 & 14,781 \\
  & Claude-3.5-Haiku & 27.5 $\pm$ 5.7 & 69.0 $\pm$ 2.3 & 32.4 $\pm$ 9.6 & 67.2 $\pm$ 10.4 & 0.0 $\pm$ 0.0 & 15,216 \\
 \midrule
 \multirow{5}{*}{w/o MLP}
  & DeepSeek-V2      & 100.0 $\pm$ 0.0 & 26.0 $\pm$ 1.0 & 100.0 $\pm$ 0.0 & 0.0 $\pm$ 0.0 & 0.0 $\pm$ 0.0 & 6,128 \\
  & Qwen             & 100.0 $\pm$ 0.0 & 26.6 $\pm$ 1.2 & 100.0 $\pm$ 0.0 & 0.0 $\pm$ 0.0 & 0.0 $\pm$ 0.0 & 5,953 \\
  & GPT-4o           & 100.0 $\pm$ 0.0 & 25.8 $\pm$ 0.6 & 100.0 $\pm$ 0.0 & 0.0 $\pm$ 0.0 & 0.0 $\pm$ 0.0 & 6,761 \\
  & GPT-4o-mini      & 100.0 $\pm$ 0.0 & 25.3 $\pm$ 0.4 & 100.0 $\pm$ 0.0 & 0.0 $\pm$ 0.0 & 0.0 $\pm$ 0.0 & 5,888 \\
  & Claude-3.5-Haiku & 100.0 $\pm$ 0.0 & 26.4 $\pm$ 1.4 & 100.0 $\pm$ 0.0 & 0.0 $\pm$ 0.0 & 0.0 $\pm$ 0.0 & 6,544 \\
 \bottomrule
 \end{tabular}
\end{table*}

To systematically evaluate the contribution of each core component in \methodname{}, we conduct an ablation study in Overcooked Plus. Specifically, we investigate the roles of three key modules:  (i) LLM;(ii) State-Action Transition Graph (Graph); and (iii) Multi-level Planning (MLP).

\textit{Ablation on LLMs}. 
To assess the impact of different LLMs, we keep the full framework intact (“Full” setting, retaining both Graph and MLP) and use different LLMs including GPT-4o, GPT-4o-mini, Qwen2-72B, DeepSeek-V2, and Claude-3.5-Haiku for comparison (other experimental settings remain identical to those in Section~\ref{sec:exp_setup}).  
The results reveal a clear performance stratification: the first-tier models (e.g., GPT-4o, DeepSeek-V2, Qwen) not only complete all quests stably but also discover all endings, including the most complex Ending~C; second-tier models (e.g., GPT-4o-mini, Claude-3.5-Haiku) can still achieve a $100\%$ completion rate under the framework’s guidance but fail to reach Ending~C entirely.  
This indicates that while the robustness of \methodname{} ensures a strong lower bound — enabling even weaker LLMs to complete basic quests — the ability to tackle the most challenging reasoning problems and explore the deepest endings is fully determined by the reasoning capability of the underlying LLM.

\textit{Ablation on Graph (w/o Graph)}.
In the “w/o Graph” setting, the system directly samples action sequences from the LLM’s outputs without constructing a state-action transition graph. As shown in Table~\ref{tab:kitchen_ablation}, removing the Graph leads to a significant performance drop: even the strongest model, GPT-4o, sees its completion rate fall from $100.0\%$ to $46.1\%$, highlighting the critical role of the Graph in ensuring stable and consistent quest execution.  
The coverage of complex endings also deteriorates markedly without the Graph, with completion rates for the most difficult Ending~C dropping sharply — in some models, to zero. This occurs because the state-action transition graph enables the merging and reuse of shared subpaths, producing a richer and more diverse set of feasible trajectories, especially for deeper or less probable endings.  
Additionally, the average steps required to complete quests increase significantly without the Graph. Lacking global knowledge of explored states and transitions, the system tends to generate longer or redundant sequences. In contrast, the Graph allows for path pruning and reuse, facilitating the generation of shorter and more efficient plans. 
These findings demonstrate that the transition graph is essential not only for increasing execution stability, but also for enabling efficient trajectory construction and deeper exploration within the solution space.

\textit{Ablation on Multi-level Planning (w/o MLP)}.
In the “w/o MLP” setting, the LLM generates entire action sequences solely based on the quest description and game instructions, which are then recorded into the state-action transition graph, with no hierarchical goal decomposition and prompt-based completion preference guidance.
The results show that all five LLMs maintain a $100.0\%$ completion rate without MLP; however, ending diversity collapses entirely, with all runs converging to the simplest Ending~A (B and C coverage: $0.0\%$). The average steps also drop significantly to around $25$–$26$. For instance, GPT-4o achieves Avg Steps / Total Steps of $25.8$ / $6{,}761$ under “w/o MLP”, compared to $66.0$ / $15{,}036$ in the full framework, with the ending diversity shifting from $A/B/C=23.2/46.4/30.4$ to $A=100.0$, $B=C=0$. DeepSeek-V2 and Qwen show the same trend.  
It is important to emphasize that this reduction in average steps does not indicate improved efficiency; instead, it reflects a shift in difficulty distribution and a metric bias caused by exclusively exploring the simplest ending. This phenomenon is analogous to the PPO results discussed in Section~\ref{sec:exp_results}.  
Therefore, the multi-level planning structure, through hierarchical goal decomposition and path guidance, effectively drives the system into longer and more complex dependency chains, thereby unlocking medium/high-difficulty branches (B/C) and improving multi-ending coverage. 

\subsection{Discussions}
Across the four benchmark environments, \methodname{} demonstrates consistently strong performance in completion reliability, ending coverage, and interaction efficiency. In Table~\ref{tab:all_game_results_with_ci}, \methodname{} achieves the highest completion rate in all four environments, reaches all designated quest endings, and uses orders-of-magnitude fewer environment interaction steps than competitive search and RL baselines. {\color{black}At the same time, the PySC2 results indicate that this advantage becomes noticeably smaller when the main challenge shifts from long-horizon dependency reasoning to real-time reactive control.}

The ablation studies further highlight the contribution of each core component. The reasoning capability of the LLM sets the upper bound of the system's performance: stronger models consistently discover more complex endings, whereas smaller-scale models, though capable of maintaining high completion rates under the framework's guidance, fail to reach high-difficulty branches. The state–action transition graph is essential for both execution stability and deep exploration; removing it leads to sharp declines in both completion rate and outcome diversity, as the agent loses the ability to reuse subpaths, prune redundant exploration, and globally coordinate trajectories. The multi-level planning structure is also critical for multi-ending coverage, as all runs collapse into the simplest ending with no MLP. 

In contrast, traditional RL methods (e.g., PPO) and search-based approaches (e.g., GA) remain fundamentally constrained by their ``search-style'' strategy paradigm, with their performance heavily reliant on the design of reward functions. These approaches suffer from the sparsity and externality of reward signals—PPO in particular tends to fall into local optima. In the Overcooked Plus, PPO often converges to the simplest ending. We hypothesize that one key bottleneck lies in the limitations of its reward function design. This issue is especially pronounced in long-horizon, high-complexity quests, where intermediate actions crucial for reaching complex endings (e.g., backtracking) may not yield immediate rewards, making it difficult for policy networks to learn them. As a result, PPO struggles to explore the full state space leading to harder endings. Crafting such fine-grained reward structures requires deep domain knowledge and extensive trial-and-error, making them difficult and costly to implement in complex game scenarios. It’s worth noting that the costs of tuning reward functions are not included in the results shown in Table~\ref{tab:all_game_results_with_ci}, and including them would further increase total overhead. ELLM-PPO, which leverages LLM-assisted intrinsic rewards, alleviates local optimum issues to some extent but is still fundamentally bound by the limitations of RL’s exploration paradigm. GA achieves slightly better diversity than PPO due to its global search nature, but at the cost of prohibitively high sample complexity and computational expense.

To address these limitations, \methodname{} takes a fundamentally different perspective, grounded in semantic-level planning rather than reward-driven search. By explicitly reasoning about the final objectives and their logical dependencies before execution, \methodname{} can generate purposeful and diverse action trajectories from a global perspective, effectively avoiding the local optima that plague RL-based methods. Furthermore, this architecture offers strong scalability: when adapting to new environments or quests, only the game mechanics, APIs, and mappings between \low{}s and game operations need to be specified—forming a knowledge-engineered pipeline—rather than retraining costly reward functions or policy networks. {\color{black}This design can potentially reduce migration and deployment costs while maintaining strong performance in long-horizon, complex quests, although the PySC2 results also reveal a clearer limitation in highly reactive settings. Prior work has suggested the potential of LLM-based online planning in PySC2~\cite{LLM-PySC2}; therefore, a natural extension of \methodname{} is to replace offline \low{} generation with online generation in highly reactive scenarios, which may improve real-time adaptation and task completion. Another promising direction is to combine \methodname{} with dedicated local planning or control algorithms, allowing the LLM to focus on narrative-level reasoning and high-level dependency management while lower-level planners handle fine-grained operational decisions. Such a hybrid design may better exploit the complementary strengths of semantic planning and reactive control in highly dynamic environments.}

{\color{black}
Beyond serving as an effective tool for initial smoke testing, our framework also has the potential to provide lasting value throughout the entire game testing lifecycle. The constructed state–action transition graph encapsulates rich knowledge of quest structures and player–environment interactions, which can be reused and extended for diverse testing activities. For instance, in regression testing, previously discovered key paths and endings can be quickly replayed to validate whether critical functionalities remain intact after game updates, reducing costs by several orders of magnitude compared to full re-exploration. In test coverage analysis, the graph enables systematic inspection of explored and unexplored branches, guiding the prioritization of future testing.

These properties also suggest a practical HCI extension of the framework. A potential interface could provide graph-based visualization of explored states and endings, editable task plans, and one-click export of validated trajectories into engineering artifacts. Such visualization could also assist quest design verification by offering designers a transparent overview of the strategy space, helping identify unintended shortcuts, dead ends, or logical flaws. While we leave formal usability evaluation to future work, the current structured outputs and transition graph already make such extensions technically feasible and well aligned with tester-centered workflows.}

\subsection{Limitations}
While the experimental results demonstrate the effectiveness and robustness of \methodname{}, several limitations remain that warrant discussion and provide directions for future improvement.

First, the multi-level planning in \methodname{} is highly dependent on the availability of quest descriptions as input. If the descriptions are ambiguous, incomplete, or do not clearly reflect the existence of multiple endings, the system’s ability to explore diverse paths can be significantly constrained. Therefore, providing high-quality and well-structured quest descriptions is a fundamental prerequisite for unlocking the full potential of \methodname{}.

Second, the quests used in this study are relatively ordinary and rely primarily on basic logic and common-sense knowledge—both of which are well-covered during the pretraining of general-purpose LLMs. This contributes to the system’s stable performance in such settings. However, when applied to more complex or domain-specific scenarios, the reasoning capability of the underlying LLM may become a bottleneck, negatively impacting overall performance. Beyond capability limitations, cost is another important consideration. Although our experiments primarily report environment interaction costs, the actual runtime cost of \methodname{} also includes the computational overhead of LLM calls. {\color{black}In other words, interaction efficiency does not fully translate to total cost efficiency. In our current evaluation, the average input token usage per quest is approximately 42k $\pm$ 9k in Overcooked Plus, 44k $\pm$ 13k in Minecraft, and 58.5k $\pm$ 20.8k in PySC2, while the corresponding average output token usage is about 7k $\pm$ 1k, 8.5k $\pm$ 1.8k, and 6.5k $\pm$ 1.8k, respectively. This inference cost is likely to increase further for games with longer quests, richer mechanics, and more complex branching structures.} {\color{black}Moreover, our preliminary experiments indicate that replacing offline \low{} generation with online generation in PySC2 could increase input-token consumption to approximately 461.2k $\pm$ 320.8k per quest, creating a clear performance--cost trade-off.} Since both performance and cost are closely tied to the quality of the base model and its alignment with the quest, future deployments must strike a balance between the two. For instance, smaller, locally deployed models fine-tuned with domain-specific data could be used to reduce inference costs while maintaining sufficient planning capability.

Third, as the complexity of games and quests increases, the state–action transition graph must be abstracted to avoid handling redundant or overly fine-grained state details, while still preserving critical nodes and connectivity. However, when the number of interactive objects and quest states grows dramatically, the graph structure can rapidly expand in size if no effective state equivalence abstraction mechanism is employed. This may lead to significant degradation in storage, retrieval, and sampling efficiency, posing further scalability challenges for the system.

\subsection{Threats to Validity}
For internal validity, the first concern lies in the inherent randomness of LLM outputs, which may affect reproducibility and performance stability. To address this, we perform each test multiple times under identical settings and report mean values with 95\% confidence intervals.  
Second, the pretraining data of LLMs is not transparent and may contain knowledge overlapping with certain quests or environments, thereby introducing a potential evaluation bias. To mitigate this, we use multiple LLMs from different vendors (e.g., GPT-4o, Qwen, DeepSeek) to reduce dependence on any single source of prior knowledge.  
Third, the performance of baseline methods may be affected by the design of reward functions and hyperparameters, while the suboptimal configurations could lead to unfair comparisons. To minimize this risk, we designed competitive reward functions and parameter settings in line with best practices.

For external validity, one key concern is the restricted scope of evaluated environments. We test on {\color{black}four} games—{\color{black}Mini RPG, Overcooked Plus, Minecraft, and a PySC2-based RTS scenario}—selected to represent diverse genres and quest structures. However, these still cannot fully reflect the wide variety of real-world games, such as FPS, puzzle, or narrative-heavy adventure games. Additional evaluations are needed to validate generalizability in future work.
Another threat concerns the realism of quest descriptions. Although we designed the quests by drawing inspiration from actual commercial games and incorporating multi-ending mechanics, the descriptions themselves are still manually constructed. Real-world game quests—especially in industrial settings—may involve noisier, less structured, or intentionally obfuscated information.
}

\section{Conclusion and Future Work}
\label{sec:conclusion}
This paper introduces a novel framework based on LLMs, \methodname{}, for automated smoke testing of multi-ending game quests. By integrating multi-level planning with a dynamic state-action transition graph, the framework effectively addresses the challenges of path exploration and efficiency in multi-ending game smoke testing.
In multi-level planning, high-level \high{} ensures broad exploration of quest paths, while low-level \low{} provides more granular planning units, enabling more precise task execution and comprehensive coverage of diverse gameplay scenarios.
The dynamic state-transition graph records the relationships between states and actions during quest execution, supporting comprehensive path coverage while eliminating path redundancy, and providing a foundation for subsequent testing.

{\color{black}In the experimental evaluation, we assess the effectiveness of the \methodname{} framework across {\color{black}four} distinct game environments: Mini RPG, Overcooked Plus, Minecraft, and PySC2. We compare \methodname{} with four baseline methods---Random, Genetic Algorithm, Proximal Policy Optimization, and ELLM-PPO.}
The results show that \methodname{} outperforms these methods in task completion rates, ending diversity, and gameplay interaction efficiency. Additionally, we conducted ablation studies to assess the contribution of multi-level planning to comprehensive exploration and the role of the graph structure in path generation. Furthermore, by comparing different LLMs, we confirmed a positive correlation between the framework's performance and the capabilities of the LLMs used.

Although the framework has shown promising results, further exploration and improvement are required in the following directions. 
First, the primary focus of future work will be to closely collaborate with industry partners to evaluate the framework’s effectiveness in real-world industrial settings. The environments used in the current experiments are still relatively simple, and evaluating the framework in actual industrial scenarios will help assess its practical applicability and ensure its relevance in more complex, real-world game development contexts.
Second, support for dynamic environments is another area for future development. The current \methodname{} framework is primarily designed for relatively static environments. However, in more complex industrial games, particularly online games, game objects and states may change unpredictably. Therefore, future work will focus on integrating real-time monitoring and adaptive planning strategies. This will ensure that the framework can maintain stable and effective testing even in highly dynamic and unpredictable environments.{
\color{black}Finally, we plan to extend \methodname{} toward a human-in-the-loop testing assistant for engineers. Since the framework already produces structured intermediate outputs and an interpretable state--action transition graph, these results can be exported in structured formats such as XML and integrated with user-facing tools. A possible interface would provide graph-based visualization of explored states and endings, editable task plans, and one-click export of validated trajectories to engineering artifacts. This direction may further improve the practical usability of the framework by helping testing engineers guide, inspect, and reuse automated testing results more effectively.
}

\section*{Acknowledgment}
This work was supported by JST SPRING (No. JPMJSP2128) and JSPS KAKENHI (No. 23K28064, 25K15290).

\bibliographystyle{IEEEtran}
\bibliography{bib}

\raggedbottom
\begin{IEEEbiography}[{\includegraphics[width=1in,height=1.25in,clip,keepaspectratio]{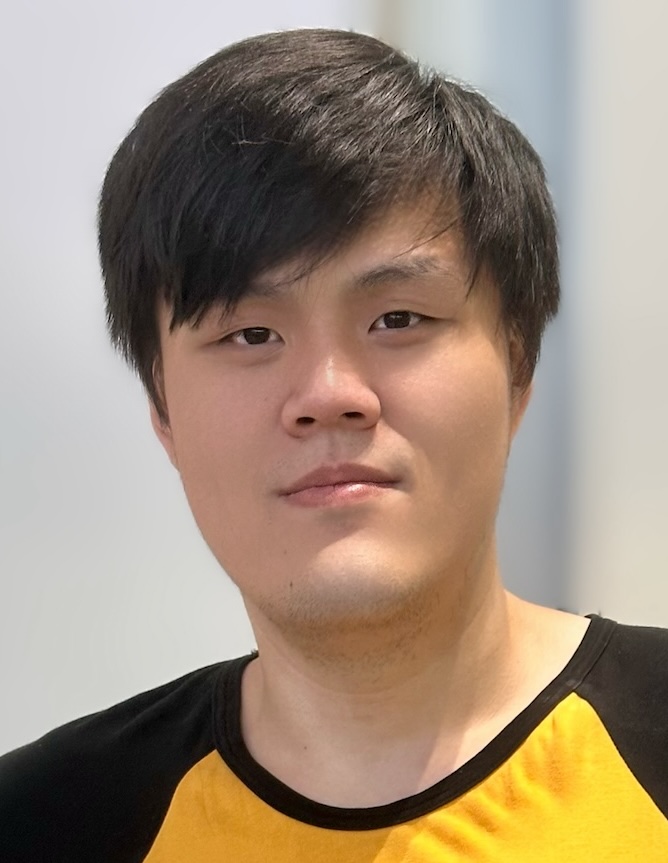}}]{Jinyu Cai}
is currently pursuing the Ph.D. degree in computer science at Waseda University, Tokyo, Japan, and is supported by JST SPRING. He received the M.E. degree in computer science from Waseda University in 2023. His research interests include autonomous agents and large language models.
\end{IEEEbiography}

\begin{IEEEbiography}[{\includegraphics[width=1in,height=1.25in,clip,keepaspectratio]{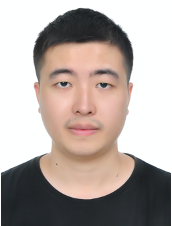}}]{Mingyue Zhang}
is an Associate Professor at Southwest University, Chongqing, China. He received the B.E. degree in computer science from Dalian University of Technology, Dalian, China, in 2017 and the Ph.D. degree in computer science from Peking University, Beijing, China, in 2023. His research interests include self-adaptive systems, multi-agent systems, and reinforcement learning.
\end{IEEEbiography}

\begin{IEEEbiography}[{\includegraphics[width=1in,height=1.25in,clip,keepaspectratio]{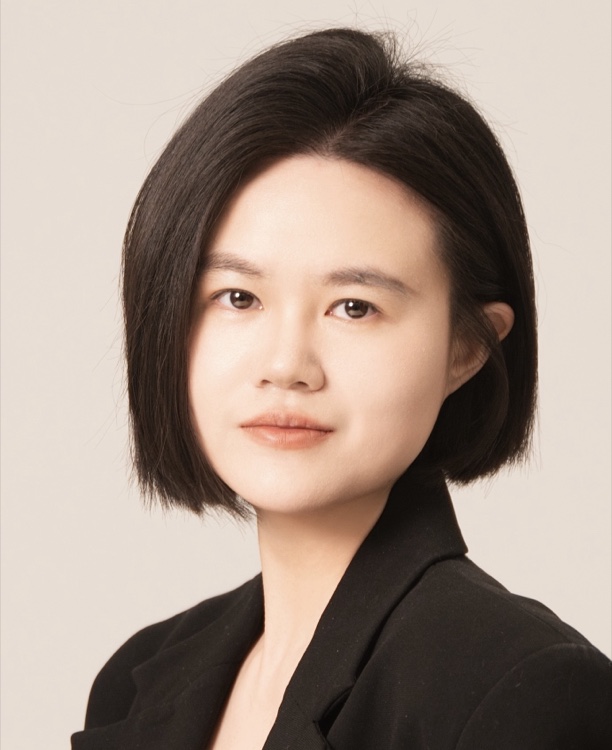}}]{Nianyu Li}
is a researcher specializing in human-involved self-adaptive systems. She received the Ph.D. degree in Computer Software and Theory from Peking University in 2021, under the supervision of Prof. Zhi Jin and Prof. Wenpin Jiao. She has also conducted research internships at the National Institute of Informatics (NII) with Prof. Zhenjiang Hu and at Carnegie Mellon University (CMU) with Prof. David Garlan. Her research interests include software modeling and analysis, requirements engineering, specification and verification, system safety, security, and cyber-physical systems.
\end{IEEEbiography}

\begin{IEEEbiography}[{\includegraphics[width=1in,clip,trim=38bp 0bp 38bp 0bp]{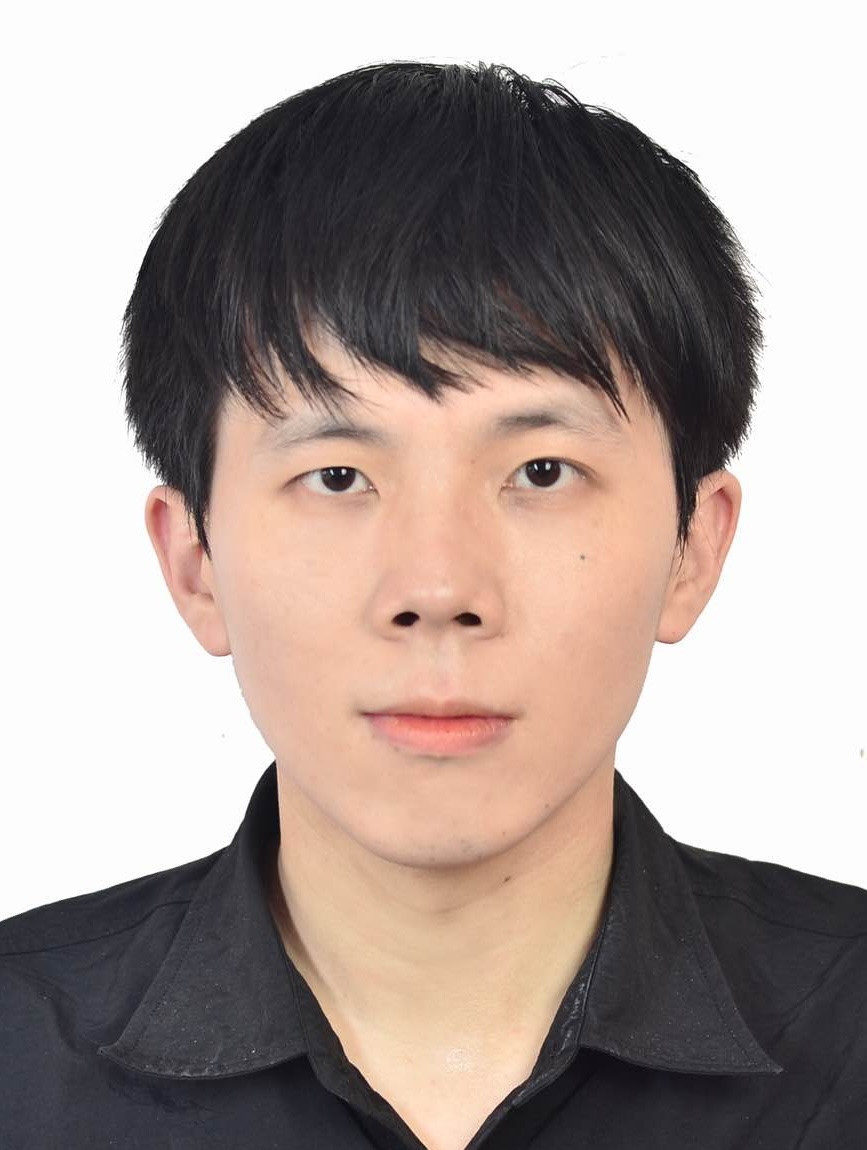}}]{Jinglue Xu}
is a Researcher at Sakana AI. He received his M.S. and Ph.D. degrees in Computer Science from the University of Tokyo in 2022 and 2025, respectively. His research interests include large language models, autonomous agents, evolutionary computation, and automated machine learning, with a particular focus on developing automated, collaborative, and accessible AI systems.
\end{IEEEbiography}

\begin{IEEEbiography}[{\includegraphics[width=1in,height=1.25in,clip,keepaspectratio]{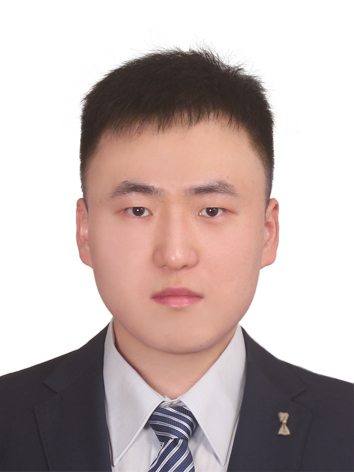}}]{Tianshui Li}
is currently pursuing the Ph.D. degree with the Department of Information and Communication Engineering, Graduate School of Information Science and Technology, the University of Tokyo, Tokyo, Japan. He received the B.S. degree in Computer Science and Engineering from the University of Aizu, Aizuwakamatsu, Japan, in 2019. His research interests include evolutionary computation, swarm intelligence, large language models, and numerical optimization.
\end{IEEEbiography}

\begin{IEEEbiography}[{\includegraphics[width=1in,height=1.25in,clip,keepaspectratio]{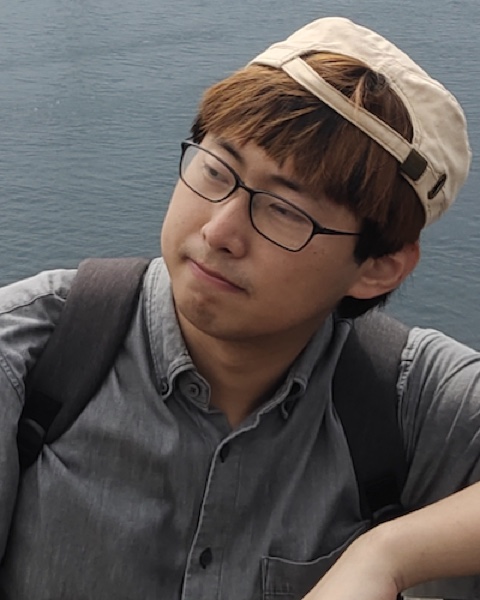}}]{Jialong Li}
is an Assistant Professor at Waseda University, Tokyo, Japan. He also serves as a Research Fellow at the Institute of Science Tokyo. He received the B.E., M.E., and Ph.D. degrees in computer science from Waseda University in 2019, 2021, and 2025, respectively. His research interests include self-adaptive systems and requirements engineering. He is a member of the IEEE, ACM, IPSJ, and IEICE.
\end{IEEEbiography}

\begin{IEEEbiography}[{\includegraphics[width=1in,height=1.25in,clip,keepaspectratio]{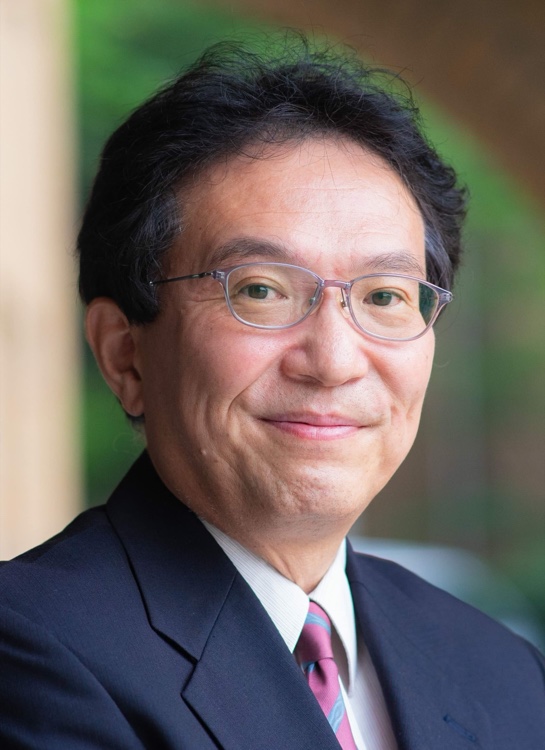}}]{Hitoshi Iba}
is a Professor with the Department of Information and Communication Engineering, Graduate School of Information Science and Technology, the University of Tokyo, Tokyo, Japan. He received the Ph.D. degree in information engineering from the University of Tokyo in 1990. He joined the Electrotechnical Laboratory in the same year and was a Visiting Researcher at Stanford University, Stanford, CA, USA, from 1996 to 1997. He joined the Department of Information and Communication Engineering, the University of Tokyo, as an Associate Professor in 1998 and became a Professor in 2004. His research interests include evolutionary systems, artificial life, and the foundations of artificial intelligence.
\end{IEEEbiography}

\begin{IEEEbiography}[{\includegraphics[width=1in,height=1.25in,clip,keepaspectratio]{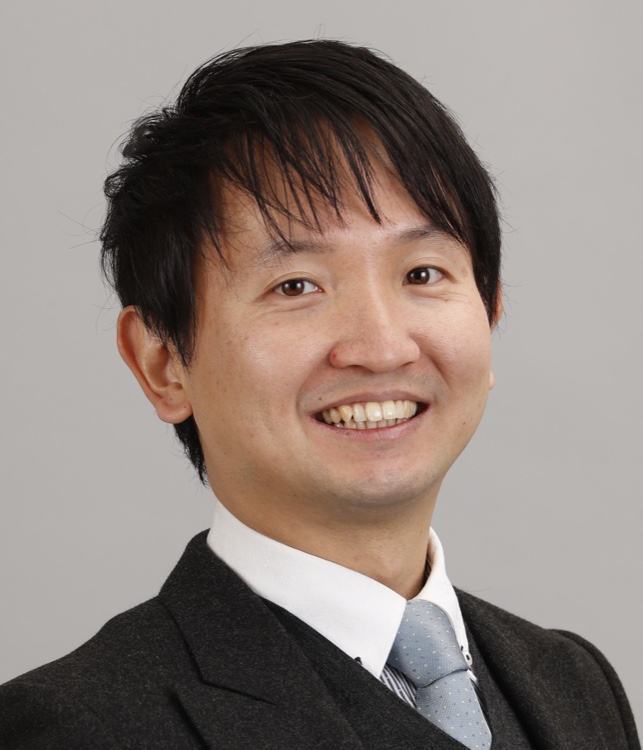}}]{Kenji Tei}
is an Associate Professor at the Institute of Science Tokyo, Tokyo, Japan. He also holds specially appointed associate professorships at Waseda University and the National Institute of Informatics, Tokyo, Japan. He received the Ph.D. degree in engineering from Waseda University in 2008. His research interests include self-adaptive systems, software architecture, and model-driven engineering. He is a member of the IEEE Computer Society, ACM, IPSJ, and IEICE, and serves as an Associate Editor of ACM Transactions on Autonomous and Adaptive Systems.
\end{IEEEbiography}

\vfill

\end{document}